\documentclass[acmtog, nonacm]{acmart}

\AtBeginDocument{%
  \fancypagestyle{plain}{%
    \fancyhf{} %
    \fancyfoot[L]{ACM SIGENERGY Energy Informatics Review}%
    \fancyfoot[R]{Volume 5, Issue 3, September 2025}}
  \providecommand\BibTeX{{%
    \normalfont B\kern-0.5em{\scshape i\kern-0.25em b}\kern-0.8em\TeX}}}
    
\let\oldmaketitle\maketitle
\renewcommand{\maketitle}{%
  \oldmaketitle%
  \thispagestyle{plain}%
  \pagestyle{plain}}

\usepackage{adjustbox}
\usepackage{graphicx}%
\usepackage{multirow}%

\usepackage{amsmath,amssymb,amsfonts}%
\usepackage{amsthm}%
\usepackage{mathrsfs}%
\usepackage[title]{appendix}%
\usepackage{xcolor}%
\usepackage{manyfoot}%
\usepackage{booktabs}%
\usepackage{listings}%
\usepackage{subcaption}
\usepackage{comment}
\usepackage{verbatim}
\usepackage{bm}
\usepackage{array}
\usepackage{textcomp}
\usepackage{stfloats}
\usepackage{url}
\usepackage{eurosym}

\usepackage{epsfig}
\usepackage{siunitx}
\usepackage{rotating}
\usepackage{tikz}
\usepackage{circuitikz}
\usepackage{wrapfig}

\newtheorem{theorem}{Theorem}%  meant for continuous numbers
\newtheorem{proposition}[theorem]{Proposition}% 
\newtheorem{definition}{Definition}%

\title{Electricity-Aware Bid Format for Coordinated Heat and Electricity Market Clearing}

\begin{document}


\author{Lesia Mitridati}
\email{lemitri@dtu.dk}
\affiliation{%
  \institution{Technical University of Denmark}
  \city{Kgs. Lyngby}
  \country{Denmark}
}

\author{Jalal Kazempour}
\email{jalal@dtu.dk}
\affiliation{%
  \institution{Technical University of Denmark}
  \city{Kgs. Lyngby}
  \country{Denmark}
}

\author{Pascal Van Hentenryck}
\email{pvh@gatech.edu}
\affiliation{%
  \institution{Georgia Institute of Technology}
  \city{Atlanta}
  \country{USA}}

%\renewcommand{\shortauthors}{Mitridati et al.}


\begin{abstract}
    Coordination between heat and electricity markets is essential to achieve a cost-effective and efficient operation of the energy system. In the current sequential market practice, the heat market is cleared before the electricity market and has no insight into the impacts of heat dispatch on the electricity market. While preserving this sequential practice, this paper introduces an electricity-aware bid format for the coordination of heat and electricity systems. This novel market mechanism defines heat bids conditionally on the day-ahead electricity prices. Prior to clearing heat and electricity markets, the proposed bid selection mechanism selects the valid bids which minimize the heat system operating cost while anticipating heat and electricity market clearing. This mechanism is modeled as a trilevel optimization problem, which we recast as a mixed-integer linear program using a lexicographic function. We use a realistic case study based on the Danish electricity and heat system and show that the proposed bid selection mechanism yields a 4.5\% reduction in the total operating cost of heat and electricity systems compared to the existing market-clearing procedure while reducing the financial losses of combined heat and power plants and heat pumps due to invalid bids by up to 20.3 million euros.
\end{abstract}

\begin{CCSXML}
<ccs2012>
<concept>
<concept_id>10010405.10010481</concept_id>
<concept_desc>Applied computing~Operations research</concept_desc>
<concept_significance>500</concept_significance>
</concept>
<concept>
<concept_id>10003752.10010070.10010099.10010106</concept_id>
<concept_desc>Theory of computation~Market equilibria</concept_desc>
<concept_significance>500</concept_significance>
</concept>
<concept>
<concept_id>10010583.10010662.10010668.10010672</concept_id>
<concept_desc>Hardware~Smart grid</concept_desc>
<concept_significance>500</concept_significance>
</concept>
</ccs2012>
\end{CCSXML}

\ccsdesc[500]{Applied computing~Operations research}
\ccsdesc[500]{Theory of computation~Market equilibria}
\ccsdesc[500]{Hardware~Smart grid}
\keywords{Sector coupling, District heating, Market design, Hierarchical optimization}


\maketitle

\section{Introduction}

%\begin{enumerate}
%	\item The participation of combined heat and power (CHP) plants, and HPs at the interface between heat and electricity systems creates strong economic and physical interdependencies between them \cite{mitridati2020appendix}.
%	\item In addition: renewables create greater volatility and uncertainty in the electricity system: which may impact the heat system too!
%	\item Current markets should interface economic and physical aspects of the systems... But physical characteristics are not accounted for... In particular networks and unit commitment constraints
%	\item This non-coordinated operation of CHPs, HPs may impact their profitability, and create non cost effective operation of the systems
%\end{enumerate}

%Pressed by growing environmental concerns, 
%The rapid growth of renewable, stochastic, and non-dispatchable energy sources has increased the need for flexibility in the electricity system \cite{lund2007renewable,lund2015review,}.

Exploiting the potential synergies between electricity and other energy systems, such as district heating, has been identified as a key solution to achieve a sustainable energy future \cite{lund2007renewable,meibom2013energy,pinson2017towards,lund2010role,lund20144th}. Indeed, a coordinated operational strategy that harnesses the flexibility of the assets at the interface between electricity and heat systems, including Combined Heat and Power (CHP) units and large-scale Heat Pumps (HPs), can facilitate the integration of additional renewable energy sources and decarbonization in both systems.
However, in several countries, such as Denmark, heat and electricity systems are operated by sequential and independent markets in the day-ahead stage \cite{Varmelast}. The main shortcomings of this market framework is that the myopic dispatch of assets at the interface of electricity and heat systems limits their operational flexibility and ability to provide ancillary services \cite{song2023potentials,fernqvist2023district}, may lead to curtailment of renewable energy sources \cite{desguers2024integration}, and risks intensifying their market power in both heat and electricity markets \cite{virasjoki2018market}.
This challenge raises an important research question: How to improve the dispatch of CHPs and HPs in the heat market in order to achieve a cost-effective and efficient operation of the overall energy system?

This question has been addressed in the literature by proposing a drastic change in the current market design, i.e., moving towards \textit{fully integrated} operational strategies and market mechanisms based on heat and electricity production co-optimization \cite{mitridati2018power,zheng2018integrated,dai2018general,maurer2023toward,babagheibi2023integrated,sorknaes2020smart}. By fully utilizing the district heating linepack storage capacity, these studies showed significant gains in terms of operating cost for the overall energy and penetration of renewable energy in the power system. However, heat and electricity markets are currently operated by independent entities, who cover vastly different geographical areas, and have different (and sometimes conflicting) objectives, timelines, and rules \cite{pinson2017towards}. Therefore, jointly dispatching heat and electricity production in an integrated market platform can only realistically be considered as an \textit{ideal benchmark}, due to the disruptive organizational and regulatory changes it would require. Additionally, while the overall operating costs may be decreased, the operating costs of one individual sector may be drastically increased, as has previously been observed in the literature \cite{mitridati2016optimal,mitridati2020heat}. Due to the aforementioned reasons, and based on extensive talks with regulators and heat market operators, we are convinced that a fully integrated market is not a feasible and realistic framework for sector coordination.

In opposition to these fully integrated operational approaches, the novel heat market-clearing procedure introduced in \cite{mitridati2020heat} claims to provide a \textit{soft coordination} between heat and electricity systems while respecting the sequential clearing of their respective markets.
%This coordination is achieved by modeling the heat market clearing as a bilevel optimization problem, which seeks to minimize the heat dispatch cost in the upper level, while anticipating the impact of heat dispatch of CHPs and HPs on the electricity market clearing in the lower level. 
This coordination is achieved by allowing the heat market clearing to model the heat production costs of CHPs and HPs as a function of day-ahead electricity prices while anticipating the impact of heat dispatch on the electricity market clearing as equilibrium constraints. Although the resulting hierarchical heat market-clearing procedure respects the sequential heat and electricity market-clearing framework, it results in increased computational complexity for the heat market operator, and pricing issues arising from non-convexities. Therefore, this approach still requires significant regulatory changes.

%Furthermore, the market-clearing procedure proposed in \cite{mitridati2020heat}  does not account for the non-convex temperature dynamics in district heating networks and the complex techno-economic characteristics of CHPs and HPs, such as their start-up and no-load costs as well as minimum on/off times. 
%To address these issues, \cite{} and \cite{} develop optimization models accounting for the dynamics of heat transfer in the district heating network, making it possible to utilize the capability of storing energy in heat pipelines. These studies illustrate the benefits of the accurate representation of techno-economic characteristics of the district heating network to increase flexibility of the overall energy system at the day-ahead stage. In addition, as CHPs and HPs are required to produce the energy traded in the day-ahead heat market regardless of electricity market outcomes, an accurate representation of their techno-economic characteristics is essential to achieve an efficient and secure operation of the heat system. However, adding such complexities requires solving a heat unit commitment, and brings pricing challenges due to introducing 0/1 binary variables \cite{ruiz2012pricing,o2005efficient}.

Other approaches in the literature have focused on enhancing bid formats and clearing mechanisms to better represent and harness the operational flexibility of assets in the markets and improve the coordination between multiple trading floors. 
Several studies introduced extended bid formats for demand response, aiming at harnessing operational flexibility from prosumers by providing a better representation of their complex preferences \cite{wang2010new,liu2015extending}. Similarly, the price-region bid format introduced in \cite{bobo2018offering} provides a unified framework to enhance the representation of the techno-economic characteristics of non-conventional flexible assets in electricity markets. In particular, this approach was applied to model the flexibility of a district-heating utility and represent its price curve as a convex function of electricity prices. Besides, the authors in \cite{byeon2019unit} introduced \textit{bid-validity} constraints in the unit commitment problem of gas-fired power plants to better account for the interdependencies between electricity and natural gas systems in their electricity bids.
The main appeal of these methods is that they remain compatible with existing market frameworks without major regulatory changes.

To tackle the above challenges, this paper makes several contributions to the state-of-the-art.
The first two contributions of this paper pertain to market design: The first contribution is to introduce an \textit{electricity-aware bid format} in the day-ahead heat market, inspired by the concept of bid-validity conditions introduced in \cite{byeon2019unit}. This format enables CHPs and HPs to design bids for the heat market, which are \textit{conditioned} on day-ahead electricity prices. Contrary to traditional bid formats, e.g., price-quantity bids, this novel bid format provides an explicit representation of the complex techno-economic characteristics of CHPs and HPs, which create interdependencies between heat and electricity markets. The second contribution is to develop an \textit{electricity-aware bid selection} mechanism which commits heat producers and selects their \textit{valid} bids, i.e., those bids that ensure \textit{cost recovery} of heat producers prior to their participation in heat and electricity markets, by anticipating the sequential clearing of heat and electricity markets. This approach does not require any major regulatory changes, as both the heat and electricity market-clearing procedures remain unchanged, and solely relies on information exchange between the heat and electricity market operators to achieve coordination. This framework is in line with recent regulatory changes, encouraging sector coordination and information exchange between market and system operators. For instance, the joint Federal Energy Regulatory Commission (FERC) and North American Electric Reliability Corporation (NERC) report on the 2011 polar vortex put an emphasis on increased information sharing between energy sectors to improve systems reliability and prevent extreme events \cite{FERCreport,FERCactions}.
The third contribution of this paper pertains to computational tractability: The proposed electricity-aware bid selection mechanism is formulated as a hierarchical (trilevel) optimization problem. We introduce a tractable reformulation as a single-level mixed-integer linear program (MILP) using a lexicographic optimization approach to represent the sequential clearing of heat and electricity markets.
Finally, the last contribution of this paper concerns a thorough numerical evaluation of the proposed mechanism on two substantial case studies. 
%This evaluation compares the proposed electricity-aware bid selection mechanism to the existing decoupled heat and electricity market mechanism, and an ideal fully integrated market mechanism. 
The first illustrative case study shows that the proposed electricity-aware bid format and bid selection mechanism can achieve $77.6\%$ of the so-called \textit{value of coordination} achieved by the fully integrated mechanism while maintaining a sequential heat and electricity market framework. The second case study, based on the realistic electricity and heat systems in Denmark, provides additional insights into how the proposed approach is able to ensure cost recovery for CHPs and HPs in the heat market while reducing the operating cost of the overall energy system by $4.5\%$ and wind curtailment by $0.4\%$ compared to the decoupled mechanism.

The remainder of this paper is organized as follows. Section \ref{section:2} provides the required preliminaries and describes the techno-economic interdependencies between heat and electricity systems and the current status-quo in the markets that operate them. Section \ref{section:3} introduces the proposed electricity-aware heat bid format and bid selection mechanism, and discusses their impacts on the heat and electricity markets. Section \ref{section:4} details the mathematical formulation of the proposed electricity-aware heat bid selection mechanism. Section \ref{section:5} presents numerical results. Finally, Section \ref{section:6} concludes and discusses potential extensions of this work.

\section{Challenges in heat and electricity markets} \label{section:2}

%Heat and electricity systems have traditionally been operated independently and sequentially. Despite the decoupled operation of heat and electricity systems, the large penetration of CHPs and HPs at the interface between the systems yields strong economic and physical interactions. In this section, we present these existing interactions and the challenges they raise for the optimal operation of heat and electricity systems. This shows the importance of accurately modeling the techno-economic characteristics of both heat and electricity systems in order to optimally coordinate their dispatch

\subsection{Existing day-ahead market framework} \label{markets_prelim}

%energy systems are operated by competitive auction-based markets which interface the physical and economic aspects of each system. 
In many European countries, including Nordic countries, the day-ahead heat and electricity markets, respectively denoted by indexes $m \in\{\text{H}, \text{E}\}$, operate on the principles of \textit{energy exchanges}. The day-ahead heat and electricity markets are currently cleared sequentially and independently, with the day-ahead heat market being cleared before the electricity market. These markets operate in similar ways: In each market $m$, a market participant $j \in \mathcal{I}_z^m$ in a given market zone $z \in \mathcal{Z}^m$ submit a set of bids $b \in \mathcal{B}^m$ for each hour of the following day $t \in \mathcal{T}$, in the form of independent non-decreasing price-quantity pairs $\left(c_{jbt}^m,\overline{s}^m_{jbt}\right)$, implicitly embedding their techno-economic characteristics. Furthermore, market participants may choose to self-commit a minimum volume of their production (or consumption) $\underline{s}^m_{jt}$, which will be dispatched regardless of the equilibrium price in their market zone. \footnote{By convention, we consider that the quantity parameters $\overline{s}^m_{jbt}$, and $\underline{s}^m_{jt}$ take positive values for generation offers and negative values for consumption bids.} Contrary to energy markets operating on the principles of \textit{energy pools}, such as North American markets, the physical flows within each market zone are neglected in the market clearing \cite{meeus2005development}. 

Two main differences exist between heat and electricity markets, resulting from the physics of the underlying networks. Firstly, while electricity markets aim at ensuring the balance of supply and demand at each time, heat markets can harness the slower dynamics in the heat network and the available linepack storage capacity to decouple supply and demand. Secondly, the scale and coupling of market zones vary significantly between the heat and power systems. In the power system, the market clearing in interconnected zones is coupled by the optimal allocation of the available transfer capacity (ATC), computed by the regional coordination centers as the expected maximum energy that can be transferred between two market zones without leading to network constraints violation within either market zone \cite{ETSO2000,tosatto2019hvdc}. This market coupling leads to zonal market prices, obtained as the marginal prices in each market zone. For instance, in the day-ahead market (operated by Nordpool), Nordic countries are split into 12 interconnected market zones, 2 in Denmark, 1 in Finland, 5 in Norway, and 4 in Sweden. On the other hand, due to slower dynamics and higher thermal losses in the heat network, district heating networks typically cover a few geographically close urban areas. Each geographically-isolated district heating network is operated by an independent market operator, and split into a single (or few) market zones. For instance, the district heating network in the greater Copenhagen area supplies over 500,000 households and the day-ahead heat market corresponds to a single market zone. Therefore, in the remainder of this paper, we consider that each heat market zone represents a single geographically-isolated district heating network.

\subsection{Interdependencies and inefficiencies}

%CHPs ($j \in \mathcal{I}^\text{CHP}$) and large-scale HPs ($j \in \mathcal{I}^\text{CHP}$) participate in both day-ahead heat and electricity markets. 
In the sequential market framework described above, CHPs and large-scale HPs participating in both heat and electricity markets create implicit interdependencies between these independently operated markets. Indeed, the techno-economic characteristics of these units link their bids in heat and electricity markets.

Firstly, the heat-driven dispatch of CHPs and HPs strongly constrains their bids in the day-ahead electricity market. Indeed, the heat $\bm{Q_{jt}}$ and electricity $\bm{P_{jt}}$ outputs of CHPs and HPs are strongly linked \cite{lahdelma2003efficient}. Indeed, HPs $j \in \mathcal{I}^\text{HP}$ produce heat from electricity at a fixed ratio $\text{COP}_j$, such that $ \bm{Q_{jt}} = - \text{COP}_j \bm{P_{jt}}$\footnote{By convention, $\bm{P_{jt}} \leq 0$ represents a power consumption.}, and CHPs $j \in \mathcal{I}^\text{CHP}$ produce heat and electricity at varying ratios, such that $r_j \bm{Q_{jt}}  \leq \bm{P_{jt}} \leq \overline{F}_j - \rho_j \bm{Q_{jt}} $, where $\overline{F}_j $ represents their maximum fuel intake, $r_j$ their minimum heat-to-power ratio, and $\rho_j$ their maximum relative heat-to-power efficiency. As a result, in the day-ahead electricity markets, CHPs and HPs must adjust the minimum $\bm{\underline{P}_{jt}}$ and maximum $\bm{\overline{P}_{jt}}$ electricity dispatch to ensure feasibility with respect to their fixed heat dispatch.
%CHPs offer their minimum and maximum electricity outputs $\bm{\underline{P}_{jt}}$, $\bm{\overline{P}_{jt}}$ at their marginal electricity production cost, and HPs bid their inflexible electricity demand $\bm{L^\textbf{HP}_{jt}}$.
In countries with high penetration of CHPs and HPs, such as Denmark, this electricity-myopic heat dispatch may limit the penetration of renewable energy sources and impact day-ahead electricity prices  \cite{mitridati2020heat}, leading to inefficiencies in both heat and electricity systems.

In addition, in order to ensure their profitability, the bids of CHPs and HPs in the day-ahead heat market must reflect their marginal heat production cost. However, the marginal production costs of CHPs and HPs are intrinsically dependent on the electricity market prices. Indeed, as HPs $j \in \mathcal{I}_z^\text{HP}$ located in electricity market zone $z \in \mathcal{Z}^\text{E}$ produce heat from electricity purchased in the day-ahead electricity market, their marginal heat production cost $\dot{\Gamma}_{jt}^{\text{H}}$ is linearly dependent on the electricity market prices $\bm{\lambda^\text{E}_{zt}}$, such that

\begin{small}
\begin{equation} \label{eq:marginal_cost_HP}
    \dot{\Gamma}_{jt}^{\text{H}} = \text{COP}_{j} \bm{\lambda^\textbf{E}_{zt}}, \forall j \in \mathcal{I}_z^\text{HP}, t\in \mathcal{T}, 
\end{equation}
\end{small}
Similarly, the variable heat production cost of CHPs $j \in \mathcal{I}_z^\text{CHP}$ located in electricity market zone $z \in \mathcal{Z}^\text{E}$ represents their total variable production costs minus revenues from electricity sales. Assuming a linear variable production\footnote{Note that this linearity assumption is without loss of generality, and the following methodology can be straightforwardly applied to any non-decreasing cost curve which can be approximated as a piecewise linear function.}, the heat production cost can be expressed as $\Gamma_{jt}^{\text{H}} = c_j (\bm{P_{jt}} + \rho_j \bm{Q_{jt}}) - \bm{\lambda^\textbf{E}_{zt}} \bm{P_{jt}}$. Therefore, their marginal heat production cost represents the incremental variable heat production cost at the optimal heat-to-power ratio, which depends on the day-ahead electricity price \cite{pinson2017towards,mitridati2016optimal,mitridati2020heat}, such that 

\begin{small}
\begin{equation} \label{eq:marginal_cost_CHP}
    \dot{\Gamma}_{jt}^{\text{H}} = \begin{cases}
        & \rho_j \bm{\lambda^\textbf{E}_{zt}} , \text{ if } \bm{\lambda^\text{E}_{zt}} \geq c_j \\
        & - r_j \bm{\lambda^\textbf{E}_{zt}} + c_j (\rho_j +r_j) , \text{ otherwise }
    \end{cases}, \forall j \in \mathcal{I}_z^\text{CHP}, t\in \mathcal{T}.
\end{equation}
\end{small}
For low electricity prices, their marginal heat production cost represents the increase in variable production cost for producing one extra unit of heat and $r_j$ extra units of electricity, and, for high electricity prices, it represents the opportunity loss for producing one extra unit of heat and $\rho_j$ fewer units of electricity. Therefore, the marginal heat production cost of CHPs can be expressed as a convex piece-wise linear function of the electricity prices.

As day-ahead electricity market prices are unknown prior to the heat market clearing, CHPs and HPs must \textit{anticipate} these prices in order to accurately compute its heat bids. In practice, each CHP and HP uses its own deterministic electricity price forecast to compute its \textit{expected} marginal heat production costs. This bid format is myopic to the impacts of the heat dispatch of CHPs and HPs on the electricity market prices, which, in turn, impact the marginal heat production costs of CHPs and HPs \cite{virasjoki2018market}. As a result, the bids of CHPs and HPs in the day-ahead heat markets may differ from their \textit{realized} marginal heat production costs due to (i) forecast errors on the day-ahead electricity market prices, and/or (ii) the exercise of market power \cite{virasjoki2018market}. This may lead to financial losses for CHPs and HPs and decreased social welfare in both heat and electricity markets. Additionally, this bid format makes heat market monitoring challenging, highlighting the need for a more efficient and transparent bid format.

\section{Towards an electricity-aware heat market} \label{section:3}

%In order to address the aforementioned challenges in the current heat and electricity markets, the following section describes the proposed electricity-aware bid format, and the resulting heat market framework which clears these bids.

\subsection{Electricity-aware bid format}

As discussed, the existing price-quantity bid format in the day-ahead heat market is unable to account for the complex interdependencies between heat and electricity markets, in particular, the interdependencies between day-ahead electricity prices and marginal heat production costs of CHPs and HPs.

One major challenge for coordinating the day-ahead heat and electricity dispatch is to respect the current sequential order and independent clearing of the markets and preserve both cost recovery and merit-order dispatch \cite{kazempour2018stochastic}. Therefore, our proposed approach is to introduce a novel electricity-aware bid format in the day-ahead heat dispatch, in the form of simple price-quantity bids $\{c^H_{jbt},s^H_{jbt}\}$, \textit{conditioned} on day-ahead electricity prices. This electricity-aware bid format will (i) improve the efficiency of heat and electricity dispatch, (ii) provide greater transparency and facilitate market monitoring, and (iii) ensure cost recovery for CHPs and HPs, which is a fundamental desirable property of the sequential and independent heat and electricity markets.

Ensuring cost recovery requires ensuring that a bid can be dispatched only if it is profitable, i.e., $c_{jbt}^H \geq \dot{\Gamma}_{jt}$. 
%, i.e. $\bm{u^\textbf{bid}_{jbt}}=0$. This can be formulated as a the following cost-recovery constraints: 
%\begin{equation} \label{eq:cost_recovery}
%    ( c_{jbt}^H - M ) \bm{u^\textbf{bid}_{jbt}} \geq \dot{\Gamma}_{jt} - M, j \in \mathcal{I}^\text{H}, b \in \mathcal{B}^\text{H}, t \in \mathcal{T},
%\end{equation}
%where $M$ is a large-enough constant. 
For a heat market participant $\forall j \in \mathcal{I}^\text{H}_z$ located in electricity market zone $z \in \mathcal{Z}^\text{E}$ whose marginal heat production cost at each time $t \in \mathcal{T}$ can be expressed as a convex piece-wise linear function of the day-ahead electricity prices, with fixed parameters $a_{jkt}$, and $b_{jkt} \in \mathbb{R}$:

\vspace{-7pt}
\begin{small}
\begin{equation}
 \dot{\Gamma}^\text{H}_{jt} = \max ( \{a_{jkt}  \bm{\lambda^\textbf{E}_{zt}} + b_{jkt} : k=1,...,K \} ),
\end{equation}
\end{small}
As a result, this cost-recovery condition can be recast as $\bm{\lambda^\text{E}} \in [\underline{\lambda}^\text{E}_{jbt},\overline{\lambda}^\text{E}_{jbt}]$,
%\begin{small}
%\begin{subequations}
%\begin{align}
%    & \bm{\lambda^\text{E}_{zt}} \leq \dfrac{c^\text{H}_{jbt} - b_{jkt}}{a_{jkt}} , \forall k \in \{1,...,K\} \text{ s.t. } a_{jkt} > 0 \\
%    & \bm{\lambda^\text{E}_{zt}} \geq \dfrac{c^\text{H}_{jbt} - b_{jkt}}{a_{jkt}} , \forall k \in \{1,...,K\} \text{ s.t. } a_{jkt} < 0 .
%\end{align}
%\end{subequations}
%\end{small}
with the bounds $\{\underline{\lambda}^\text{E}_{jbt},\overline{\lambda}^\text{E}_{jbt}\}$ within which this bid is profitable defined as

\begin{small}
 \begin{subequations}
\begin{align}
    & \underline{\lambda}^\text{E}_{jbt} = \max ( \{ \dfrac{c^\text{H}_{jbt} - b_{jkt}}{a_{jkt}} :  k =1,...,K \text{ s.t. } a_{jkt} < 0 \} \cup \{\underline{\lambda}^\text{E}_z\} ) \\
    & \overline{\lambda}^\text{E}_{jbt} = \min ( \{ \dfrac{c^\text{H}_{jbt} - b_{jkt}}{a_{jkt}} :  k = 1,...,K \text{ s.t. } a_{jkt} < 0 \} \cup \{\overline{\lambda}^\text{E}_z\} ) , 
\end{align}
\end{subequations}   
\end{small}
and $\underline{\lambda}^\text{E}_z$, $\overline{\lambda}^\text{E}_z$ are the minimum and maximum electricity bid prices allowed in this market zone.

%%%% definition of bid format
As the marginal heat production cost of CHPs and HPs can be formulated as convex piece-wise linear functions of electricity prices, as seen in \eqref{eq:marginal_cost_HP}-\eqref{eq:marginal_cost_CHP}, to ensure cost-recovery, their price-quantity bids in the heat market $(c_{jbt}^H,s_{jbt}^H)$ should be defined over a range of electricity prices $\{\underline{\lambda}^E_{jbt},\overline{\lambda}^E_{jbt}\}$ over which this bid is profitable. This new electricity-aware bid format is defined formally below. %In this approach, each heat price-quantity bid in the heat market is associated with bounds $\{\underline{\lambda}^E_{jbt},\overline{\lambda}^E_{jbt}\}$ on day-ahead electricity prices for which the bid is considered valid.
\begin{definition}[Electricity-aware heat bid format]
For each heat market participant $j \in \mathcal{I}^\text{H}$ and  time step $t \in \mathcal{T}$, an electricity-aware heat bid $b \in \mathcal{B}^\text{H}$ is defined as a price-quantity pair $\{c^H_{jbt},s^H_{jbt}\}$ associated with a range $\{\underline{\lambda}^E_{jbt},\overline{\lambda}^E_{jbt}\}$ of electricity prices over which this bid is considered valid and can be dispatched.
\end{definition}

%where $M_j$ is a big-enough constant, which provides an upper bound on the marginal heat production costs of CHPs and HPs. Decision variables $\bm{u^\textbf{bid}_{jbt}} \in \{0,1\}$ ensure that $\bm{u^\textbf{bid}_{jbt}}=1$ if and only if bid $b$ is selected in the heat market. Therefore, the \textit{cost-recovery condition} \eqref{eq:bid_validity} guarantees that a necessary condition for a bid $b$ to be selected is that its bidding price $c^\text{H}_{jbt}$ is greater than or equal to the marginal heat production cost of this unit, i.e., $\dot{\Gamma}^\text{H}_{jt} $. Note that, for $\bm{u^\textbf{bid}_{jbt}}=0$, \eqref{eq:bid_validity} simply enforces the upper bound, i.e., $M_j \geq \dot{\Gamma}^\text{H}_{jt}$. In practice, this condition ensures cost recovery of CHPs and HPs in the heat market, conditionally on future electricity prices.

\subsection{Bid selection mechanism: A hierarchical optimization formulation}

We introduce an electricity-aware bid selection mechanism for the proposed electricity-aware bid format. 
%This mechanism aims at selecting the \textit{valid} bids submitted by all heat market participants, which minimize the operating cost of the overall heat system while anticipating their impact on the participation of CHPs and HPs in the sequential heat and electricity markets. 
This mechanism is designed to identify and select the \textit{valid} bids, i.e. those ensuring cost recovery, submitted by heat market participants, with the goal of minimizing the total operating cost of the heat system. The validity of these bids is computed endogenously based on their anticipated influence on the participation of CHPs and HPs in the sequential heat and electricity markets. This electricity-aware bid selection mechanism is implemented by the heat system operator by enforcing the following linear \textit{bid-validity conditions}:

\begin{small}
\begin{subequations} \label{eq:elec_aware_condition}
 \begin{align}
    & \bm{\lambda^E_{zt}} - \overline{\lambda}^E_{jbt} \leq M \left( 1 - \bm{u^\textbf{bid}_{jbt}} \right) , \forall z \in \mathcal{Z}^\text{E}, j \in \mathcal{I}^\text{H}_z, t \in \mathcal{T}, b \in \mathcal{B}^\text{H},  \\
    &  \underline{\lambda}^E_{jbt} - \bm{\lambda^E_{zt}} \leq M \left( 1 - \bm{u^\textbf{bid}_{jbt}} \right) , \forall z \in \mathcal{Z}^\text{E}, j \in \mathcal{I}^\text{H}_z, t \in \mathcal{T}, b \in \mathcal{B}^\text{H},
 \end{align}
\end{subequations}
\end{small}
with $M$ a large-enough constant. These conditions enforce that a bid can only be selected, i.e., $\bm{u^\textbf{bid}_{jbt}}=1$, if electricity prices are within the bounds $\{\underline{\lambda}^E_{jbt},\overline{\lambda}^E_{jbt}\}$. These bounds are directly computed by each heat market participant based on their marginal heat production costs, such that cost recovery is guaranteed within this range of prices, i.e., $\dot{\Gamma}^\text{H}_{jt} \geq c^\text{H}_{jbt}$.
%\begin{equation} \label{eq:bid_validity}
%\left(c^\text{H}_{jbt} - M\right) \bm{u^\textbf{bid}_{jbt}} \geq \dot{\Gamma}^\text{H}_{jt} - M, \forall j \in \mathcal{I}^\text{H}, t \in \mathcal{T}, b \in \mathcal{B}^\text{H}.
%\end{equation} 
%In particular, for CHPs and HPs whose marginal heat production costs can be expressed as an affine or convex piece-wise linear functions of the day-ahead electricity prices such as CHPs and HPs, the detailed expression of these electricity price bounds is provided in the online appendix.

\begin{figure}[ht]
	\centering
	\includegraphics[width=0.99\linewidth]{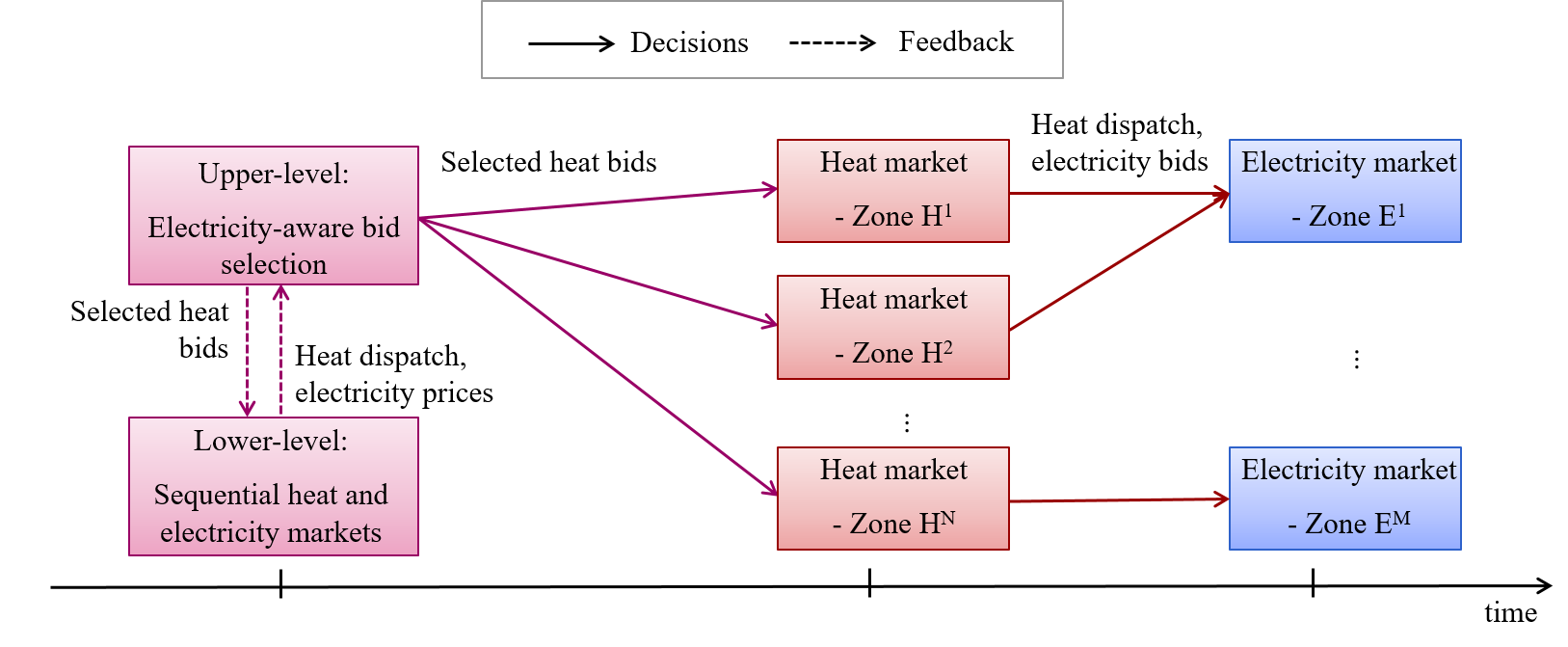}
	\caption{Sequential bid selection mechanism and sequential clearing of heat and electricity markets}
	\label{fig:EABS}
\end{figure}

Once the bid selection mechanism has selected the valid bids, CHPs and HPs can participate in the day-ahead heat market by solely submitting their valid bids, in the form of independent price-quantity pairs. Once their heat dispatch is fixed in the heat market, CHPs and HPs can then participate in the day-ahead electricity market, in which they adjust their minimum and maximum electricity outputs. This sequence of bid selection and market clearing is highlighted in Fig. \ref{fig:EABS}. As isolated heat networks may interface with the same electricity market zones, this electricity-aware bid selection mechanism must coordinate the bids of CHPs and HPs across multiple heat market zones. Therefore, the proposed mechanism is solved centrally, as illustrated in Fig. \ref{fig:EABS}. As a result, this electricity-aware mechanism coordinates the participation of these units in both heat and electricity markets while the heat and electricity market-clearing mechanisms remain unchanged.
%Additionally, the proposed heat unit commitment is able to model the techno-economic characteristics of CHPs and HPs and anticipate the impact of their commitment on both heat and electricity market clearings through electricity-aware heat bids. 

The proposed bid selection mechanism is built upon a Stackelberg game with a single leader and multiple followers. The leader is the heat system operator solving a bid selection problem that determines the validity of the bids of CHPs and HPs by anticipating the reaction of the followers, i.e., the sequential heat and electricity market clearings. The action of the leader constrains the reaction of the followers, and, in return, the reaction of the followers impacts the objective function of the leader. In our case, the selected bids affect the bids submitted in the heat market. In return, the dispatch of CHPs and HPs in the heat market affects the operating cost of the heat system in the objective of the bid selection mechanism, and the day-ahead electricity prices impact the bid-validity constraints.

Under the assumption of perfect information on the input data of the heat and electricity markets, our proposed bid selection mechanism can be formulated as a hierarchical optimization problem, as illustrated in Fig. \ref{fig:EABS}. The upper-level problem represents the heat bid selection mechanism, which is constrained by an embedded (lower-level) optimization problem, representing the sequential and independent clearing of the heat and electricity markets, in which the heat market is cleared first. The detailed formulation of these optimization problems is provided in Section \ref{section:4}.

The proposed electricity-aware heat bid format provides CHPs and HPs with greater flexibility to take into account their techno-economic characteristics and the linkage between their heat and electricity outputs in the current sequential market design. In addition, the proposed approach can be implemented in the current sequential market framework without major regulatory and organizational changes by introducing a bid selection mechanism prior to the clearing of sequential heat and electricity markets, as described in this section.

\subsection{Impact on heat and electricity markets}

In the proposed electricity-aware market framework, the heat and electricity market clearing mechanisms remain unchanged after the bid selection step, and, thus, retain two fundamental desirable economic properties of the existing heat and electricity markets with price-quantity bids and uniform pricing, namely \textit{budget balance} for the market operators, and \textit{cost recovery} for all market participants \cite{myerson1981optimal}. This important feature guarantees that the proposed electricity-aware bid format can be implemented in the current sequential market framework without major changes. In addition, the proposed electricity-aware heat market mechanism strengthens the cost-recovery property for CHPs and HPs. Indeed, the existing heat market mechanism can only guarantee cost recovery with respect to the CHPs and HPs' submitted price-quantity bids, representing their \textit{expected} marginal production costs. However, under the assumption of perfect information, the proposed electricity-aware heat market mechanism guarantees cost recovery for CHPs and HPs with respect to their \textit{realized} marginal production costs, by discarding the electricity-aware bids that would be invalid given the realized electricity prices. This mechanism relies on the exchange of information on the electricity market supply and demand curves between the electricity and the heat system operators. Yet, if perfect information is not available, additional privacy-preserving mechanisms for Stackelberg games have been developed that can be straightforwardly applied to this mechanism, and provide guarantees on the sub-optimality of the solutions \cite{fioretto2019ppsm,mitridati2022differentially}.

In addition, similarly to the existing heat and electricity market clearing mechanisms, the proposed electricity-aware bid selection mechanism can only guarantee \textit{incentive compatibility} and \textit{market efficiency} under the assumption of perfect competition \cite{wilson1977bidding,hobbs2003complementarity}.
%Otherwise, individual market participants may increase their own profits and reduce social welfare by submitting strategic bids in both heat and electricity markets. 
Yet, \cite{virasjoki2018market} showed that, in the existing market framework, CHPs may exercise significant market power by i) strategically withholding capacity in the electricity market to increase electricity prices and revenues from electricity sales; and, in anticipation, ii) lowering their bid price in the heat market to displace production from heat-only units, in turn, lowering the opportunity loss of capacity withholding in the electricity market. Since the proposed bid-selection mechanism endogenously represents the dependence between the heat dispatch of CHPs and electricity prices, it may reject these invalid bids from CHPs, which, inherently discourages this cross-market strategic behavior. 

Finally, since information on electricity market supply and demand curves is \textit{solely} communicated to the heat market operators, and not the market participants, the proposed bid format does not raise additional \textit{privacy issues} for electricity market participants, or provide any additional strategic advantage to CHPs and HPs.

\section{Mathematical formulation and solution method} \label{section:4}

%This section introduces the mathematical formulation and a tractable reformulation of the proposed bid selection mechanism. The proofs of all propositions are provided in the Appendix.

\subsection{Upper level: Bid selection mechanism}

The proposed bid selection mechanism in the upper-level problem seeks to minimize the operating cost of the overall heat system for each hour of the following day $t \in \mathcal{T}$ while anticipating the impact of its decisions on the heat and electricity market-clearings. The set of decision variables $\Omega^\text{UL}$ of this problem includes the commitment state $\bm{u_{jt}^0}$, start-up $\bm{v^\textbf{SU}_{jt}}$, and shut-down $\bm{v^\textbf{SD}_{jt}}$ states of all heat market participants, the validity $\bm{u^\textbf{bid}_{jbt}}$ of all heat bids, defined as binary variables as binary variables \eqref{ul6}, as well as the dispatch $\bm{Q_{jbt}}$ of all heat bids and the electricity market prices $\bm{\lambda_{zt}^{\textbf{E}}}$ in all electricity market zones.
%This model can formulated as follows:
%The commitment $u^0_{jt}$ of CHPs and HPs impact their dispatch $Q_{jbt}$ in the day-ahead heat market.
%The dispatch of CHPs and HPs impact the electricity production of CHPs and electricity consumption of HPs in the electricity market, and therefore the electricity locational marginal prices (prices) $\lambda^\text{E}_{zt}$.
It can be formulated as a hierarchical optimization problem as follows:

\begin{small}
\begin{subequations} \label{ul}
	\begin{alignat}{2}
	& \min_{\Omega^\text{UL}} \ && \Theta^\text{UL} = \sum_{t \in \mathcal{T}} \sum_{j \in \mathcal{I}^{\text{H}}} \Big( c_{jt}^0 \bm{u_{jt}^0}    +  c^\textbf{SU}_{j}\bm{v^\textbf{SU}_{jt}} + \sum_{k =1}^{B^\text{H}} c_{jbt}^\text{H} \bm{Q_{jbt}}  \Big) \label{ul1} \\  
	%& \text{s.t.} \ &&	\bm{c^\textbf{SU}_{jt}} \geq  c_{jh}^\uparrow \big(   \bm{u_{jt}^0} - \sum_{k = t - h}^{t} \bm{u_{jk}^0} \big) , \forall j \in \mathcal{I}^{\text{H}} , t \in \mathcal{T}\setminus \Phi^{u,\text{init}}_{j} , h \in \Phi^r_{jt} \label{ul2}  \\	
	%%
	%& \quad && \bm{c^\textbf{SU}_{jt}} \geq 0  , \forall j \in \mathcal{I}^{\text{H}} , t \in \mathcal{T} \label{ul3} \\
	%%
	%& \quad && \bm{u^0_{jt}} = u_{jt}^{\text{init}} , \forall j \in \mathcal{I}^\text{H} , t \in \Phi^{u,\text{init}}_{j} \label{ul4} \\
	%%
	%& \quad && \sum_{k=t-\underline{\Phi}^\uparrow_{j} + 1}^{t} \bm{v^\uparrow_{jk}} \leq \bm{u_{jt}^0} , \forall j \in \mathcal{I}^{\text{H}} , t \in \mathcal{T} \setminus \Phi^{u,\text{init}}_{j} \label{ul5} \\
	%%
	%& \quad && \sum_{k=t-\underline{\Phi}^\downarrow_{j} + 1}^{t} \bm{v^\uparrow_{jk}} \leq 1 - \bm{u^0_{j(t-\underline{\Phi}^\downarrow_{j})}}    , \forall j \in \mathcal{I}^{\text{H}} ,  t \in \mathcal{T} \setminus \Phi^{u,\text{init}}_{j} \label{ul6} \\
	%%
    & \text{s.t. } && \text{bid-validity constraints } \eqref{eq:elec_aware_condition} \label{ul5} \\
	& \quad && \bm{u^\textbf{bid}_{jbt}} \leq \bm{u^\textbf{bid}_{j(b-1)t}}  , \forall j \in \mathcal{I}^{\text{H}} ,  b \in \mathcal{B}^\text{H}\setminus \{1\} , t \in \mathcal{T} \label{ul3} \\
 	& \quad && \bm{u^\textbf{bid}_{j1t}} \leq \bm{u^0_{jt}}  , \forall j \in \mathcal{I}^{\text{H}} , t \in \mathcal{T} \label{ul4} 
    \end{alignat}
    \begin{alignat}{2}
	%%
	%& \quad && \bm{\Phi_{z\tilde{z}t}} = \sum_{k=0}^{\overline{\Phi}} k \bm{u^\Phi_{z\tilde{z}tk}} , \forall z \in \mathcal{Z}^\text{H}, \tilde{z} \in \mathcal{Z}_z^\text{H} , t \in \mathcal{T} \label{ul9} \\
	%%
	%& \quad && \sum_{k=0}^{\overline{\Phi}} \bm{u^\Phi_{z\tilde{z}tk}} = 1 , \forall z \in \mathcal{Z}^\text{H}, \tilde{z} \in \mathcal{Z}_z^\text{H} , t \in \mathcal{T} \label{ul10} \\
	%%
	%%
	%& \quad && \bm{\overrightarrow{u}_{z\tilde{z}t}} + \bm{\overrightarrow{u}_{\tilde{z}zt}} = 1 , \forall z \in \mathcal{Z}^\text{H}, \tilde{z} \in \mathcal{Z}_z^\text{H} , t \in \mathcal{T} \label{ul11} \\
	%%
	%%
	%& \quad &&   \bm{\overrightarrow{u}_{z\tilde{z}t}}, \bm{u^{\Phi}_{z\tilde{z}tk}} \in \{0,1\} , \forall  z \in \mathcal{Z}^\text{H}, \tilde{z} \in \mathcal{Z}_z^\text{H} , k \in \{0,...,\overline{\Phi}\} , t \in \mathcal{T} \label{ul13}  \\
 	& \quad && \bm{v^\textbf{SU}_{jt}}  , \bm{v^\textbf{SD}_{jt}}  , \bm{u_{jt}^0} , \bm{u^\textbf{bid}_{jbt}} \in \{0,1\} , \forall  j \in \mathcal{I}^{\text{H}} , b \in \mathcal{B}^\text{H} , t \in \mathcal{T} \label{ul6} \\
	%%%%
    & \quad && \bm{v^\textbf{SU}_{jt}}  - \bm{v^\textbf{SD}_{jt}}  = \bm{u^0_{jt}}  - \bm{u^0_{j(t-1)}} , \forall j \in \mathcal{I}^{\text{H}}, t \in \mathcal{T} \label{ul2} \\
    %%%%
    & \quad && \bm{v^\textbf{SU}_{j1}}  - \bm{v^\textbf{SD}_{j1}}  = \bm{u^0_{j1}}  - u^\textit{init}_{j} , \forall j \in \mathcal{I}^{\text{H}} \label{ul2-0} \\
    %%%%%%
	& \quad && \{ \bm{Q_{jbt}} , \bm{\lambda^\textbf{E}_{zt}} \} \in \text{primal and dual solutions of market clearings \eqref{lex}} \label{ul7}.
	\end{alignat}
\end{subequations}
\end{small}
The heat system operating cost in \eqref{ul1} includes start-up costs $c^\textbf{SU}_{jt}$, no-load costs $c_{jt}^0$, and the cost of dispatching these heat bids $c_{jbt}^\text{H}$ in the heat market. In addition, the selected bids must satisfy feasibility and consistency requirements: i) Eq. \eqref{ul5} ensures that only \textit{valid} bids are selected; and ii) Since the submitted bid prices $c^{H}_{jbt}$ are non-decreasing, \eqref{ul3}-\eqref{ul4} enforces that a bid $b$ can only be chosen if all preceding bids $1$,...,$b-1$ have been selected and the unit has been committed. Finally, the logical relationship governing the commitment, start-up, and shut-down statuses of each unit and their initial commitment status $u^\textit{init}_j$ is enforced in \eqref{ul2}-\eqref{ul2-0}.
%Constraints \eqref{ul9.1} and \eqref{ul9.2} define the discrete time delays $\bm{\Phi_{z\tilde{z}t}}$ between the market zones of the heat network, such that $\bm{\Phi_{z\tilde{z}t}} = k $ if and only if the binary variable $\bm{u^\Phi_{z\tilde{z}tk}} = 1$. %Constraint \eqref{ul10} ensures that the first mass flow entering a pipeline is the first mass flow exiting it. 
%Constraint \eqref{ul11} states the relationship between the direction of the heat flows between two connected market zones of the district heating network, where $\bm{\overrightarrow{u}_{z\tilde{z}t}} = 1$ if and only if the flow entering the pipeline at time step $t$ is directed from market zone $z$ to market zone $\tilde{z}$. 
%%%%%
Finally, the heat dispatch $\bm{Q_{jbt}}$ and electricity market price $\bm{\lambda_{zt}^{\textbf{E}}}$ variables are defined endogenously as the primal solution of the heat market clearing and the dual solution of the electricity market clearing \eqref{ul7}.

\subsection{Lower-level: Sequential heat and electricity markets}

The sequential and independent clearing of the day-ahead heat and electricity markets is embedded as a lower level of the bid validity mechanisms \eqref{ul7}, in which the heat market is cleared first.

These two market-clearing problems are formulated below.

\subsubsection{Day-ahead heat market problem:}

For given values of the commitment $\bm{u^0_{jt}}$ of CHPs and HPs and selected bids $\bm{u^\textbf{bid}_{jbt}}$, the day-ahead heat market aims at minimizing the heat dispatch cost across all interconnected market zones
%{\color{black} As the heat networks are isolated, the heat market-clearing problems in the respective market zones are independent. Therefore, multiple independent market-clearing problems can be modeled as a single optimization problem in the middle-level without the loss of generality.}
% while enforcing maximum heat transfer capacities, for given values of the time delays in pipelines fixed in the upper-level problem. This results in a linear formulation of the heat market clearing in the middle-level problem, as detailed in  Appendix. 
over the set of decision variables $\Omega^\text{H}$ including the dispatch of all submitted bids $\bm{Q_{jbt}}$, as well as heat flows $\bm{f^{\textbf{H}}_{z\tilde{z}t}}$ between market zones of the district heating network. It can be formulated as a linear optimization problem as follows:

\begin{small}
\begin{subequations} \label{heat}
	\begin{alignat}{2}
	& \min_{\Omega^\text{H}} \ && \Theta^\text{H}(\bm{Q_{jbt}}) = \sum_{t \in \mathcal{T}} \sum_{j \in \mathcal{I}^{\text{H}}} \sum_{b \in \mathcal{B}^\text{H}} c_{jbt}^\text{H} \bm{Q_{jbt}} \label{heat1} \\
	& \text{s.t.} \ &&  \sum_{j \in \mathcal{I}_z^{H}} \bm{Q_{jt}} = L^\text{H}_{zt} + \sum_{\tilde{z} \in \mathcal{Z}_z^{H}} \bm{f^{\textbf{H}}_{z\tilde{z}t}}, \forall  z \in \mathcal{Z}^\text{H}, t \in \mathcal{T} \label{heat2} \\
	& \quad && \bm{Q_{jt}} = \sum_{b \in \mathcal{B}^\text{H}} \bm{Q_{jbt}} , \forall  j \in \mathcal{I}^\text{H},t \in \mathcal{T} \label{heat3} \\
    & \quad && \underline{s}^\text{H}_{j} \bm{u^0_{jt}} \leq \bm{Q_{jt}}  , \forall  j \in \mathcal{I}^\text{H},t \in \mathcal{T} \label{heat4} \\
	%%
	%& \quad && \overline{s}^\text{H}_{jbt} \bm{u^\textbf{bid}_{j(b-1)t}} \leq \bm{Q_{jbt}} \leq \overline{s}^\text{H}_{jbt} \bm{u^\textbf{bid}_{jbt}} , \forall  j \in \mathcal{I}^\text{H}, t \in \mathcal{T} , b \in \mathcal{B}^\text{H} \label{heat5} \\
    & \quad && 0 \leq \bm{Q_{jbt}} \leq \overline{s}^\text{H}_{jbt} \bm{u^\textbf{bid}_{jbt}}, \forall  j \in \mathcal{I}^\text{H}, t \in \mathcal{T} , b \in \mathcal{B}^\text{H}\label{heat6} \\
 %%
	%& \quad &&  \bm{Q_{jb_{B^\text{H}}t}} \leq \overline{s}^\text{H}_{jb_{B^\text{H}}t} \bm{u^\textbf{bid}_{jb_{B^\text{H}}t}} , \forall  j \in \mathcal{I}^\text{H}, t \in \mathcal{T} \label{ml5.2} \\
	%%
	& \quad &&  - \underline{f}^\text{H}_{z\tilde{z}t} \leq \bm{f^{\textbf{H}}_{z\tilde{z}t}} \leq \overline{f}^\text{H}_{z\tilde{z}t}, \forall z \in \mathcal{Z}^\text{H}, \tilde{z} \in \mathcal{Z}_z^\text{H}, t \in \mathcal{T} \label{heat7} \\
    & \quad && \bm{f^\textbf{H}_{z\tilde{z}t}} = - \bm{f^\textbf{H}_{\tilde{z}zt}}, \forall z \in \mathcal{Z}^\text{H}, \tilde{z} \in \mathcal{Z}_z^\text{H}, t \in \mathcal{T}.\label{heat8}
	\end{alignat}
\end{subequations}
\end{small}
The objective function \eqref{heat1} represents the dispatch cost of all (valid) heat bids. Constraint \eqref{heat2} enforces the heat balance in each heat market zone, \eqref{heat3} defines the total heat production $\bm{Q_{jt}}$ of each market participant as the sum of its bids dispatched, \eqref{heat4} sets lower bounds on the heat production of all market participants based on their minimum capacity $\underline{s}^\text{H}_{j}$ and commitment, \eqref{heat6} sets upper bounds on the dispatch of all heat bids based on the bid quantity $\overline{s}^\text{H}_{jbt}$, \footnote{As the bids submitted are required to be non-decreasing, a bid $b$ cannot be dispatched if previous bids are not fully dispatched.} and such that only valid bids selected in the upper-level \eqref{ul} can be dispatched,
%This constraint imposes that if one of the following bids $\tilde{b}>b$ is selected, bid $b$ should fully be dispatched. 
\eqref{heat7} sets upper and lower bounds ($\overline{f}^\text{H}_{z\tilde{z}t}$, $\underline{f}^\text{H}_{z\tilde{z}t}$) on the heat flow between heat market zones, and \eqref{heat8} links the directed heat flows between market zones.

\subsubsection{Day-ahead electricity market:}

Once the heat market has been cleared, the day-ahead electricity market is cleared for the fixed values of the heat dispatch of CHPs and HPs $\bm{Q_{jt}}$. It is assumed that for any value of the heat dispatch, the electricity market clearing problem is feasible and bounded. This market clearing problem aims at minimizing the electricity dispatch cost, based on the set of decision variables $\Omega^\text{E}$ including the dispatch $\bm{P_{jbt}}$ of all electricity bids, and the power flow $\bm{f^\textbf{E}_{z\tilde{z}t}}$ between electricity market zones. It can be formulated as a linear optimization problem as follows: 

\begin{small}
\begin{subequations} \label{elec}
	\begin{alignat}{2}
	& \min_{\Omega^\text{E}} \ && \Theta^\text{E}(\bm{P_{jbt}}) = \sum_{j \in \mathcal{I}^{\text{E}}} \sum_{b \in \mathcal{B}^\text{E}} c_{jbt}^\text{E} \bm{P_{jbt}} \label{elec1} \\
 & \text{s.t.} && \sum_{j \in \mathcal{I}_z^\text{E}} \bm{P_{jt}}  = L^\text{E}_{zt} + \sum_{\tilde{z} \in \mathcal{Z}_z^\text{E}} \bm{f^\textbf{E}_{z\tilde{z}t}} , \forall z \in \mathcal{Z}^\text{E},t \in \mathcal{T}  \label{elec2} \\
	& \quad &&  \bm{P_{jt}}  = \sum_{b \in \mathcal{B}^\text{E}} \bm{P_{jbt}} , \forall j \in \mathcal{I}^\text{E},t \in \mathcal{T}  \label{elec3} \\
	& \quad &&  \bm{\underline{P}_{jt}} \leq \bm{P_{jt}} \leq \bm{\overline{P}_{jt}} , \forall j \in \mathcal{I}^\text{E},t \in \mathcal{T}  \label{elec4} \\
    & \quad && 0 \leq  \bm{P_{jbt}} \leq  \overline{s}^\text{E}_{jbt} , \forall j \in \mathcal{I}^\text{E}, b \in \mathcal{B}^\text{E} , t \in \mathcal{T} \label{elec5} \\
	& \quad && - \underline{f}^\text{E}_{z\tilde{z}t} \leq \bm{f^\textbf{E}_{\tilde{z}zt}} \leq \overline{f}^\text{E}_{z\tilde{z}t} , \forall z \in \mathcal{Z}^\text{E}, \tilde{z} \in \mathcal{Z}_z^\text{E} t \in \mathcal{T}  \label{elec6} \\
	& \quad && \bm{f^\textbf{E}_{z\tilde{z}t}} = - \bm{f^\textbf{E}_{\tilde{z}zt}}, \forall z \in \mathcal{Z}^\text{E}, \tilde{z} \in \mathcal{Z}_z^\text{E}, t \in \mathcal{T}
	\label{elec7} \\
   & \quad && \bm{\overline{P}_{jt}} = \begin{cases} & \sum_{b \in \mathcal{B}^\text{E}}\overline{s}^\text{E}_{jbt} , \forall j \in \mathcal{I}^\text{E} \setminus \mathcal{I}^\text{H} , t \in \mathcal{T} \\
   & - \dfrac{\bm{Q_{jt}}}{\text{COP}_j} , \forall j \in \mathcal{I}^\text{HP} , t \in \mathcal{T} \\
   & \overline{F}_j - \rho_j \bm{Q_{jt}} , \forall j \in \mathcal{I}^\text{CHP} , t \in \mathcal{T} \end{cases} \\
   & \quad && \bm{\underline{P}_{jt}} = \begin{cases}
   & \underline{s}^\text{E}_{jt} , \forall j \in \mathcal{I}^\text{E} \setminus \mathcal{I}^\text{H} , t \in \mathcal{T} \\
   & - \dfrac{\bm{Q_{jt}}}{\text{COP}_j} , \forall j \in \mathcal{I}^\text{HP} , t \in \mathcal{T} \\
   & r_j \bm{Q_{jt}}, \forall j \in \mathcal{I}^\text{CHP} , t \in \mathcal{T} \end{cases} \label{elec8}
	\end{alignat}
\end{subequations}
\end{small}
The objective function \eqref{elec1} represents the dispatch cost of all electricity bids. Constraint \eqref{elec2} enforces power balance in each electricity market zone, \eqref{elec3} defines the total electricity production $\bm{P_{jt}}$ of each market participant as the sum of its bids dispatched, \eqref{elec4} sets upper and lower bounds ($\bm{\overline{P}_{jt}}$, $\bm{\underline{P}_{jt}}$) on the self-committed electricity dispatch of all market participants, \eqref{elec5} sets upper bounds $\overline{s}^\text{E}_{jbt}$ on the dispatch of all electricity bids, \eqref{elec6} sets upper and lower bounds ($\overline{f}^\text{E}_{z\tilde{z}t}$, $\underline{f}^\text{E}_{z\tilde{z}t}$) on the power exchanged between electricity market zones, and \eqref{elec7} links the directed power exchanged between these market zones. Additionally, \eqref{elec8} defines the upper and lower bounds on the electricity dispatch as their maximum bids $\sum_{b \in \mathcal{B}^\text{E}}\overline{s}^\text{E}_{jbt}$ and self-committed power $\underline{s}^\text{E}_{jt}$ for conventional electricity market participants, and as functions of their heat dispatch (fixed in the heat market-clearing problem \eqref{heat}) for CHPs and HPs. Finally, the electricity zonal prices $\bm{\lambda_{zt}^\textbf{E}}$ are defined as the dual variables of the balance equations \eqref{elec2}.

%The authors in \cite{pineda2016capacity} proposed an interesting alternative that models these sequential market-clearing problem in the lower-level problem as a bilevel optimization problem, whose leader is the electricity market-clearing problem, and whose follower is the heat market-clearing problem. This alternative is illustrated in Fig. \ref{fig:3models}(b). With this setup, the electricity market-clearing problem becomes the middle-level problem of the electricity-aware bid selection mechanism, while the heat market-clearing problem becomes the lower-level problem. Despite its computational tractability, this alternative has its own shortcoming. These equilibrium (Fig. \ref{fig:3models}(a)) and bilevel (Fig. \ref{fig:3models}(b)) formulations are not equivalent. In the bilevel formulation proposed in \cite{pineda2016capacity}, the electricity market in the middle level is able to anticipate the outcomes of the heat market in the lower level, and, in case of multiple solutions to the lower-level problem, to choose the optimistic solution which minimizes its own objective. Due to the existing sequential order of these market-clearing problems, this formulation is not realistic.

\subsubsection{Lexicographic optimization formulation:}

Different approaches have been studied in the literature to model the sequential and interdependent clearing of energy markets \cite{pineda2016capacity}. 
Sequential market clearings have traditionally been modeled as an equilibrium problem, in which the second market-clearing problem takes as input the optimal solutions of the first one. Using this formulation would result in both the heat \eqref{heat} and electricity \eqref{elec} market-clearing problems as lower-level problems for the bid-selection mechanism in the upper-level \eqref{ul}. However, the dual formulation of these two lower-level problems contains non-convex bilinear terms, making the solution of the bid selection mechanism computationally challenging. The authors in \cite{pineda2016capacity} proposed an interesting alternative that models sequential market-clearing problems as a bilevel optimization problem. Using this approach would require modeling the heat market-clearing \eqref{heat} as the lower level of the electricity market-clearing \eqref{elec}. Despite its computational tractability, this alternative has its own shortcomings and is not equivalent to the equilibrium formulation. Indeed, in the case of multiple solutions, this approach chooses an optimistic solution which minimizes the electricity market costs. Due to the existing sequential order of these market-clearing problems, this formulation is not realistic.

In order to offer a computationally tractable and realistic formulation of this sequential market clearing, we observe that, for a fixed solution of the bid selection mechanism, the heat market clearing \eqref{heat} is cleared first and does not depend on the outcomes of the electricity market clearing. Therefore, its objective is optimized in priority, regardless of the solutions of the electricity market clearing. As a result, and in order to circumvent the limitations of the aforementioned approaches, we propose a novel formulation of the sequential heat and electricity market clearings as a linear lexicographic optimization problem, such that:

\begin{small}
\begin{subequations} \label{lex}
	\begin{alignat}{2}
    & \min_{\Omega^\text{H}\cup\Omega^\text{E}} \ &&  < \Theta^\text{H}(\bm{Q_{jbt}}) ,  \Theta^\text{E}(\bm{P_{jbt}}) > \label{lex1} \\
 & \text{s.t. } && \text{Eqs. } \eqref{heat2} - \eqref{heat8} ' \quad \eqref{elec2} - \eqref{elec8}.  \label{lex2}
	\end{alignat}
\end{subequations}
\end{small}
The aim of this multi-objective optimization problem is to minimize the heat and electricity dispatch costs, ranked in a lexicographic order \eqref{lex1}, subject to the primal constraints of the heat and electricity market clearings \eqref{lex2}. In this approach, one first minimizes the heat dispatch cost $\Theta^\text{H}(\bm{Q_{jbt}})$, then holding $\Theta^\text{H}(\bm{Q_{jbt}})$
constant, minimizes the electricity dispatch cost $\Theta^\text{E}(\bm{Q_{jbt}})$.
\begin{proposition}\label{prop1}
Any optimal solution to the proposed lexicographic optimization formulation \eqref{lex} of the sequential heat and electricity market clearing is an optimal solution to the equilibrium problem between the heat \eqref{heat} and electricity \eqref{elec} market clearing problems.
\end{proposition}
Note that the proposed lexicographic formulation gives priority to the heat market clearing due to the sequential order in which the heat and electricity markets are currently cleared in most Nordic countries.

\subsection{Reformulation as Mixed Integer Linear Program (MILP)}

%a third alternative, based on . which is also a bilevel optimization model. However, the middle-level problem represents the heat market clearing, which is constrained by the electricity market clearing in the lower level, as illustrated in Fig. \ref{fig:3models}(c).
%Therefore, the electricity market in the lower level is constrained by the solutions of the heat market in the middle level.

The proposed bid selection mechanism \eqref{ul} can be formulated in a compact manner as 

\begin{small}
\begin{subequations} \label{ul_compact}
	\begin{alignat}{2} 
    &  \underset{\bm{z} \in \{0,1\}, \bm{x^\textbf{H}} \geq 0, \bm{y^\textbf{E}} }{\min} \ && c^{\text{bid}^\top} \bm{z} + c^{\text{H}^\top} \bm{x^\textbf{H}} \label{ul_compact1} \\
    & \text{ s.t. } && \bm{z} \in \mathcal{Z}^{\text{bid}} \label{ul_compact2} \\
    & \quad && A^\text{bid }\bm{z} +  B^\text{bid }\bm{y^\textbf{E}} \geq b^\text{bid} \label{ul_compact3} \\
    & \quad && \{\bm{x^\textbf{H}} , \bm{y^\textbf{E}} \} \in \text{ primal and dual solutions of: } \\
    & \quad && \quad \quad \quad \quad 
    \underset{\bm{x^\textbf{H}} , \bm{x^\textbf{E}}  \geq \bm{0}}{\min} \ < c^{\text{H}^\top}   \bm{x^\textbf{H}} , c^{\text{E}^\top}   \bm{x^\textbf{E}} > \label{ul_compact4} \\
    &  \quad && \quad \quad \quad \quad \quad \text{ s.t. } \quad A^{\text{H}}  	 \bm{x^\textbf{H}} + B^{\text{H}}   \bm{z} \geq b^{\text{H}} \label{ul_compact5} \\
    & \quad  && \quad \ \quad \ \quad \ \quad \quad \quad \quad  A^{\text{E}}  \bm{x^\textbf{E}}   +  B^{\text{E}}  \bm{x^\textbf{H}} \geq b^{\text{E}} \label{ul_compact6}
	\end{alignat}
\end{subequations}
\end{small}
where $\bm{z}$, $\bm{x^\textbf{H}}$ and $\bm{x^\textbf{E}}$ represent the vectors of primal variables of the bid selection mechanism, heat and electricity market clearings, respectively, and $\bm{x^\textbf{E}}$ is the vector of dual variables of the electricity market clearing. The expression of the vectors $c^\text{bid}$, $c^{\text{H}}$, $b^{\text{H}}$, $c^{\text{E}}$, $b^{\text{bid}}$, $b^{\text{H}}$, $b^{\text{E}}$ and matrices $A^{\text{bid}}$, $B^{\text{bid}}$, $A^{\text{H}}$, $B^{\text{H}}$, $A^{\text{E}}$, $B^{\text{E}}$ can be derived from the detailed formulations of \eqref{ul}-\eqref{elec}.

\begin{proposition} \label{prop2}
The bid selection mechanism in \eqref{ul_compact} can be asymptotically approximated by the following single-level MILP\footnote{the bilinear terms in \eqref{primal_dual8} can be linearized using an exact McCormick relaxation \cite{mccormick1976computability}}:

\begin{small}
\begin{subequations} \label{primal_dual}
	\begin{alignat}{7}
& \min_{\overset{\bm{z}\in\{0,1\}^N,\bm{x^\textbf{H}} \geq \bm{0} }{\underset{\bm{x^\textbf{E}} \geq \bm{0} , \bm{y^\textbf{H}},\bm{y}^\textbf{E}}{}}} &&  \gamma c^{\text{bid}^\top}  \bm{z}  + \gamma c^{\text{H}^\top}  \bm{x^\textbf{H}} + \left( 1-\gamma \right) c^{\text{E}^\top}  \bm{x^\textbf{E}} \label{primal_dual1} \\
& \quad \quad \text{s.t.} &&  \bm{z} \in \mathcal{Z}^{\text{bid}} \label{primal_dual2} \\
& \quad &&  A^{\text{bid}}  \bm{z}  +  \dfrac{1}{(1-\gamma)} B^{\text{bid}} \bm{y^\textbf{E}}   \geq  b^{\text{bid}} \label{primal_dual3} \\
& \quad       && \eqref{ul_compact5} - \eqref{ul_compact6} \label{primal_dual5} \\
& \quad       && \bm{y^{\textbf{H}^\top}}  A^{\text{H}}  + \bm{y^{\textbf{E}^\top}} B^{\text{E}}   \leq \gamma c^{\text{H}^\top} \label{primal_dual6} \\
 &  \quad    &&  \bm{y^{\textbf{E}^\top}} A^{\text{E}}  \leq \left( 1-\gamma \right) c^{\text{E}^\top} \label{primal_dual7} \\
 & \quad &&  \bm{y^{\textbf{H}^\top}}   \left( b^{\text{H}} - B^{\text{H}}    \bm{z}  \right)  + \bm{y^{\textbf{E}^\top}} b^{\text{E}} \geq \gamma c^{\text{H}^\top}  \bm{x^\textbf{H}} + \left( 1-\gamma \right) c^{\text{E}^\top}  \bm{x^\textbf{E}}.  \label{primal_dual8} 
	\end{alignat}
\end{subequations} 
\end{small}
When the penalty factor $\gamma$ tends to 1, the solutions of \eqref{primal_dual} converge to the solutions of \eqref{ul_compact}.
\end{proposition}
This linear reformulation and the proof of Proposition \ref{prop2} are detailed in Appendix.

\section{Numerical analysis} \label{section:5}

\subsection{Case study setup} 

\begin{figure}[htbp]
    \centering
    \begin{subfigure}[b]{0.45\textwidth}
        \includegraphics[width=\textwidth]{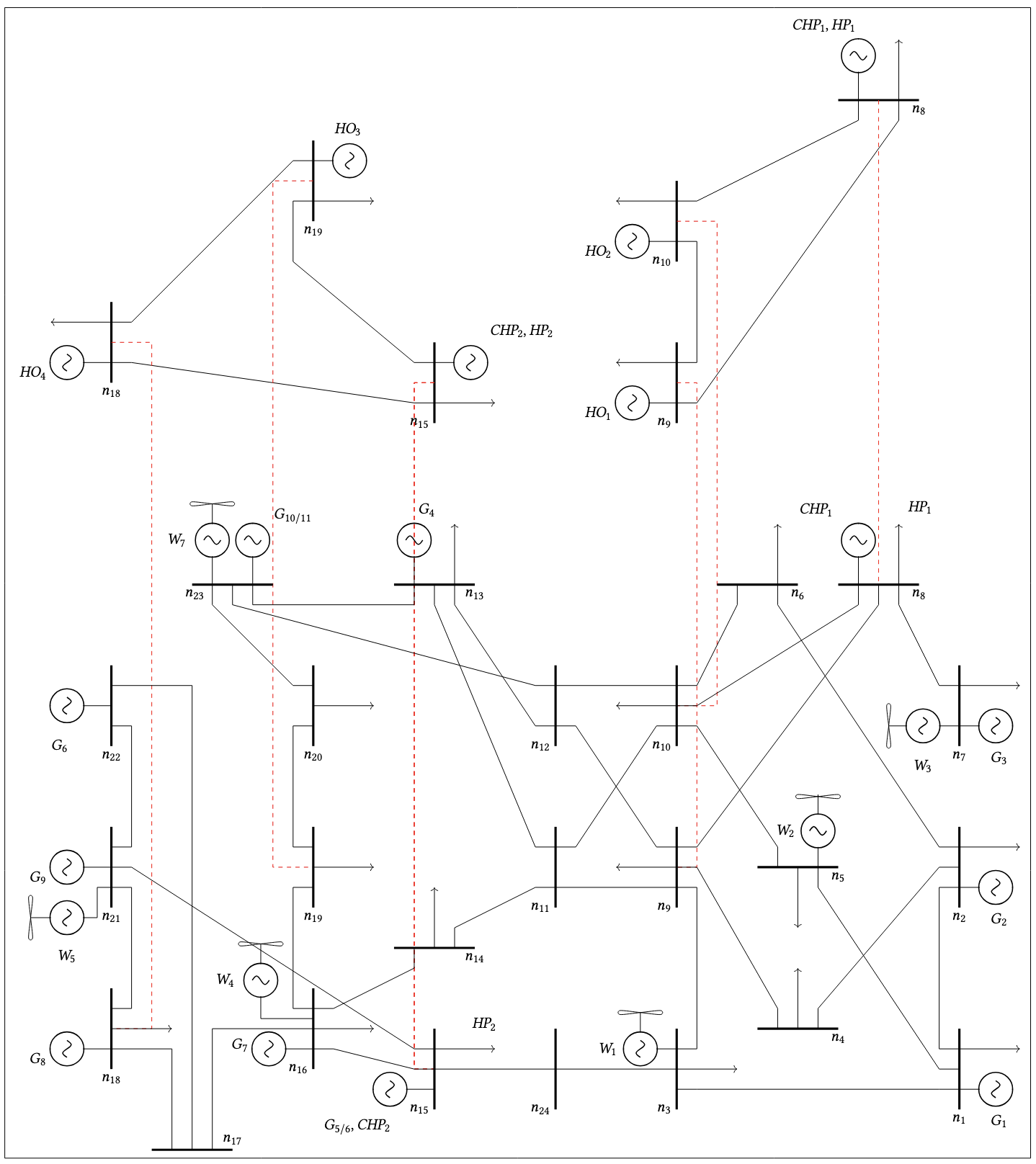}
        \caption{Case study 1: Modified IEEE 24-node electricity system with 6 wind farms (the bottom system) connected to two isolated 3-node district heating systems (the two systems on the top of the figure)}
        \label{fig:IEEE}
    \end{subfigure}
    \hfill
    \begin{subfigure}[b]{0.45\textwidth}
        \includegraphics[width=\textwidth]{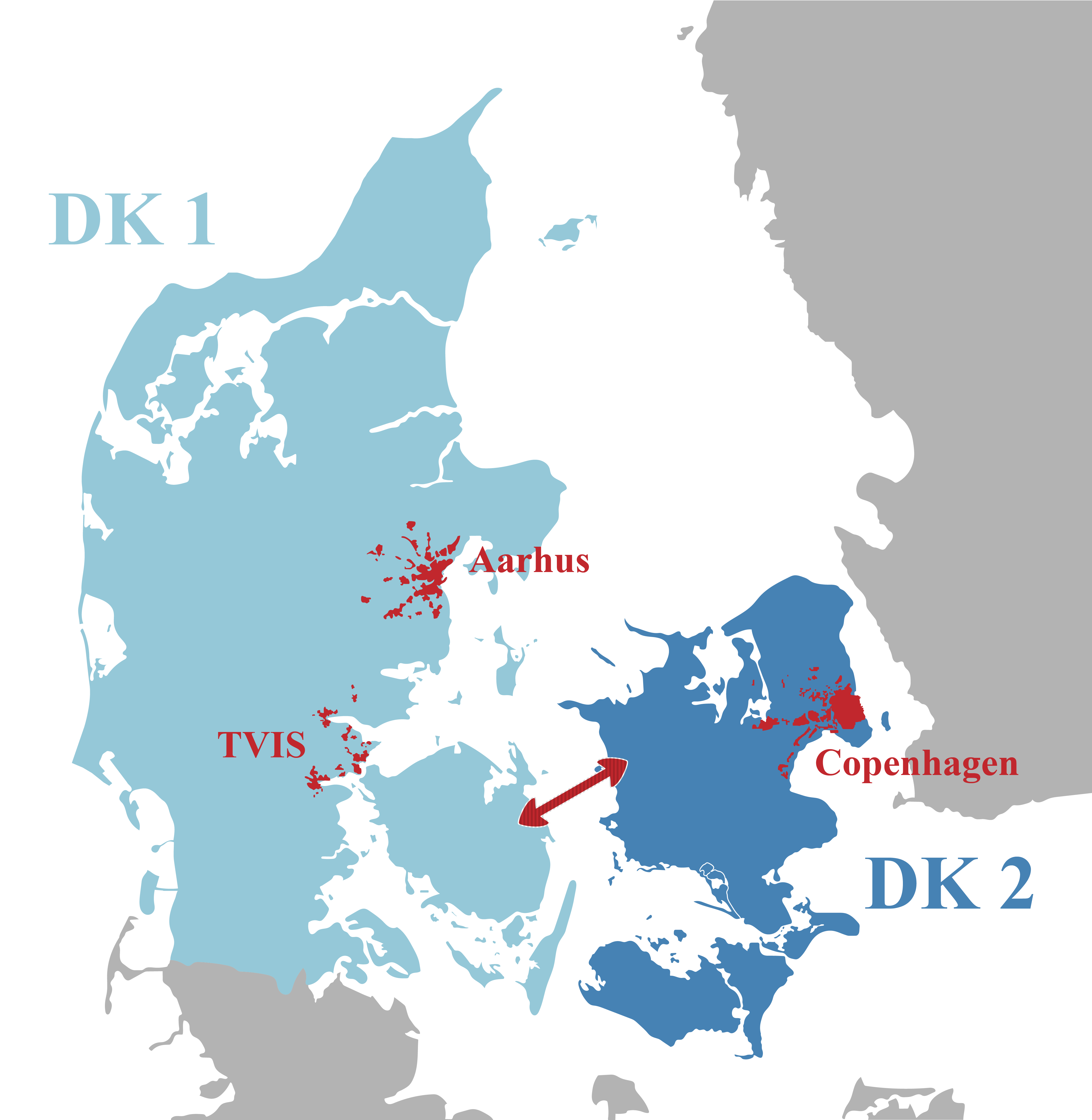}
        	\caption{Case study 2: Danish electricity market zones ($DK1$ and $DK2$) and three district heating networks (Aarhus, Copenhagen, TVIS). Red zones represent areas with high density of district heating consumption. Red arrow represents the interconnection between the two electricity market zones.}
	\label{fig:denmark}
    \end{subfigure}
    
    \caption{Heat and power networks in two case studies: Modified IEEE 24-bus system (left) and Danish energy system (right).}
    \label{fig:combined_case_studies}
\end{figure}

This section analyzes the benefits of the proposed \textit{electricity-aware} bid selection mechanism in terms of renewable energy penetration, cost-effectiveness, and profitability of CHPs and HPs, through two case studies. The first case study, representative of the greater Copenhagen area, is a modified 24-bus IEEE Reliability Test System connected to two isolated 3-node district heating networks, comprising  12 thermal power plants (G), 6 wind farms (W), 2 extraction CHPs, 2 HPs, and 2 waste incinerator heat-only units (HO), as illustrated in Fig. \ref{fig:IEEE}. Data for heat and power generation and costs, loads and wind scenarios over 2 months of operation is derived from \cite{ordoudis2016updated,zugno2016commitment,li2016combined,mitridati2018power}. This small test case represents an energy system with a particularly high penetration of wind production, as well as a large share of CHPs and HPs, which can provide operational flexibility at the interface between heat and electricity systems. As illustrated in Fig. \ref{fig:denmark}, the second case study, comprising 11 CHPs, 6 incinerators (IS), i.e., CHPs with fixed heat-electricity ratio, 6 HPs, 20 HO units and peak boilers, and $3$ heat storage tanks (HS), is representative of the Danish energy system. The electricity network is divided into two interconnected market zones, $DK1$ and $DK2$, and the district heating network is divided into three isolated markets zones. Generation costs and parameters, ATCs, loads, as well as wind and solar power generation over 1 year of operation is derived from \cite{energinet,tosatto2019hvdc,Madsen2015,mitridati2018power}. In both case studies, the penalty factor $\gamma$ in \label{milp} is fixed to $0.99$. A sensitivity analysis reveals that solutions are stable around this value, and, therefore, are assumed to have converged. 

The proposed electricity-aware bid selection mechanism (EA) is compared to: (a) the existing sequential and decoupled day-ahead market mechanism (Dec) described in Section \ref{section:2}; and (b) the fully integrated heat and electricity unit commitment and market mechanism (Int), which jointly and simultaneously optimizes the dispatch and flows in the heat and electricity systems over the following 24 hours. This integrated mechanism is not realistic in practice and is solely used as an ideal benchmark\footnote{Note that the integrated mechanism is not simulated in Case Study 2 because, due to the intertemporal constraints, when cleared sequentially day-by-day over 1 year, it is not guaranteed to result in lower operational costs compared to other mechanisms, and therefore, it does not provide a relevant benchmark.}. In order to ensure a fair comparison between the electricity-aware and decoupled mechanisms, the hourly heat bids of CHPs and HPs are computed each day as their heat marginal costs using day-ahead electricity price forecasts, as presented in Section \ref{section:2}. These forecast prices are derived by jointly clearing the heat and electricity markets.

%Details on these case studies setup and all relevant data are provided in the online appendices, available at \cite{online_appendix_1,online_appendix_2}.

\subsection{Case study 1: Modified 24-bus IEEE Reliability Test System}

This first analysis focuses on quantifying the \textit{value of coordination} achieved by the proposed electricity-aware bid selection mechanism. As summarized in Table \ref{table:costs}, the integrated market mechanism reduces the overall system cost by $11.3\%$ compared to the decoupled mechanism. The absolute cost difference, i.e., $1,330$\euro, between the decoupled and integrated market mechanisms represents the value of coordination between heat and electricity markets \cite{mitridati2018power}. However, as previously discussed, this integrated mechanism is not a realistic alternative to a sequential heat and electricity market framework. Indeed, it can be observed that the heat system operating cost is drastically increased with the integrated mechanism compared to the decoupled and hierarchical mechanisms, which supports the findings in \cite{mitridati2016optimal,mitridati2020heat}. Meanwhile, the proposed electricity-aware mechanism achieves $77.6\%$ of this so-called value of coordination while maintaining the sequential and independent structure of the markets.

\begin{table}[ht]
\centering
	\caption{Case study 1: Total, heat and electricity system costs, in $10^3$\euro, and value of coordination achieved, in \% of total value of coordination for the decoupled, electricity-aware, and fully integrated market frameworks.}
\begin{adjustbox}{width=\columnwidth,center}
	\label{table:costs}
		\begin{tabular}{lccc}
			\hline
			& \textbf{Decoupled} & \textbf{Electricity-Aware} & \textbf{Integrated} \\
			\textbf{Total cost}  & $11.75$ & $10.72$ & $10.42$ \\
			\textbf{Heat cost} &  $2.66$ & $2.96$ & $10.23$ \\
			\textbf{Electricity cost} &  $9.09$ & $7.76$ & $0.19$ \\
			%\textbf{Coordination value} & $0$ & $77.6$ & $100$ \\
			\hline
	\end{tabular}
\end{adjustbox}
  \end{table}
  
%The proposed electricity-aware heat dispatch achieves these benefits by better anticipating the impact of CHPs and HPs on the electricity market, and exploiting the operational flexibility at the interface between heat and electricity systems, as will further be discussed in Case study 2.

%As illustrated in Fig. \ref{fig:Q_1}(b) for a specific day, by anticipating the impact of the commitment and dispatch of CHPs and HPs on the electricity market prices, the proposed electricity-aware mechanism rejects the bids of $\text{CHP}_1$ during these hours, and switches heat production to the waste incinerators $\text{HO}_1$. As illustrated in Fig. \ref{fig:Q_1}(a), during these hours, the decoupled market framework accepts a lower bid, which decreases the heat system cost. However, this myopic heat dispatch results in day-ahead electricity prices which do not support cost-recovery for $\text{CHP}_1$. This illustrative example shows how the proposed bid-validity bids can provide a better representation of the techno-economic characteristics of CHPs and HPs and exploit their flexibility at the interface between heat and electricity systems.

\subsection{Case study 2: Danish energy system}

%Both mechanisms, i.e.,  decoupled and electricity-aware, are solved daily over 1 year. Each day, the hourly heat bids and no-load costs of CHPs, incinerators, and HPs are computed using the forecast day-ahead electricity prices in the market zones of the electricity system to which they are connected, as discussed in Section \ref{section:2}. 
%Based on these bids, the heat unit commitment, followed by the heat and then electricity markets are cleared for each hour of the following day. The commitment and energy storage capacity at the end of the day, are used to initialize the commitment and dispatch decisions the following day.
 
\begin{figure}[ht]
	\centering
	\includegraphics[width=\linewidth]{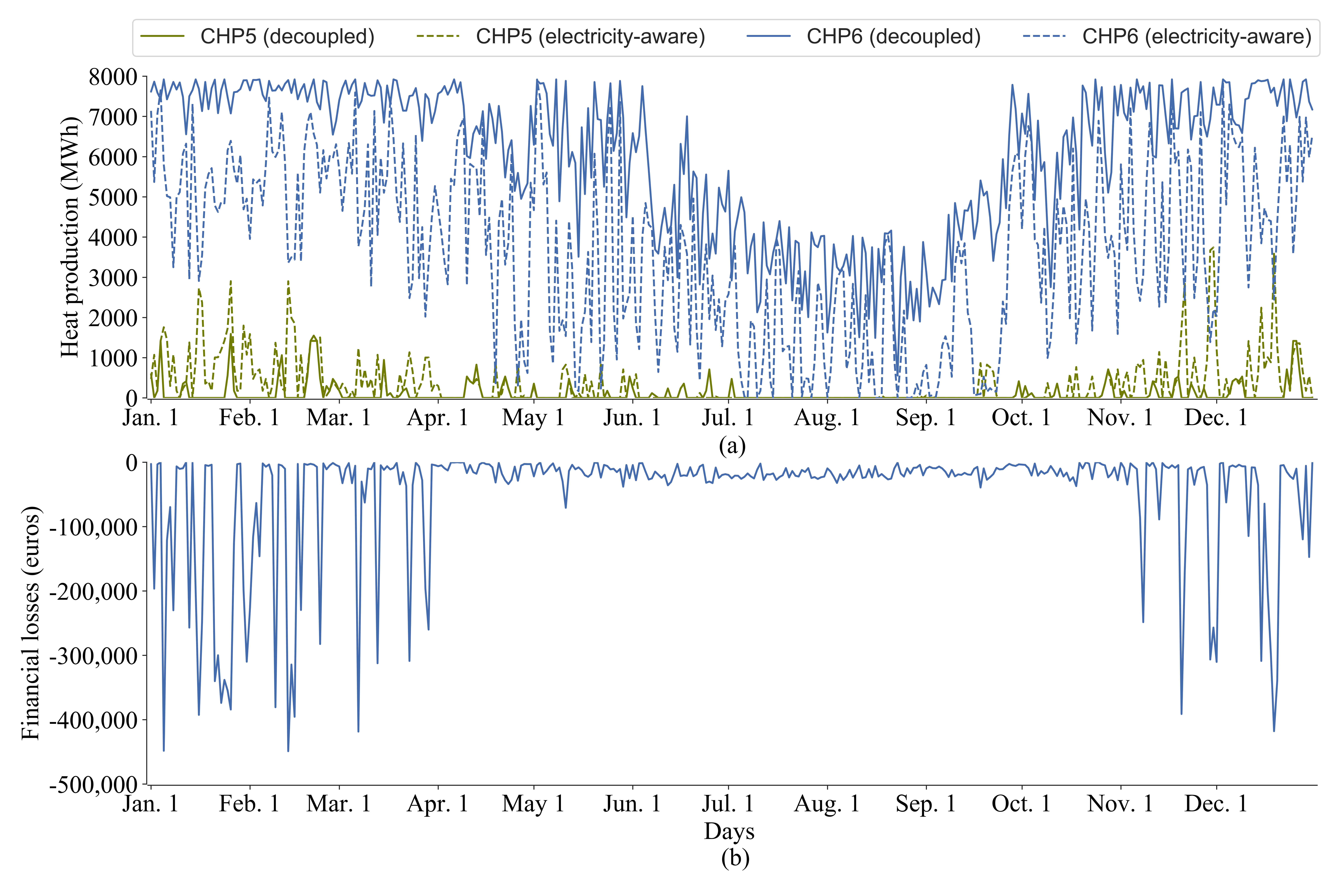}
	\caption{Case study 2: (a) Daily heat dispatch (in MWh) of CHPs with decoupled and electricity-aware mechanisms over 1 year, and (b) corresponding financial losses (in \euro) with the decoupled mechanism.}
	\label{fig:loss}
\end{figure}

This second analysis provides further insights into the operation of the heat and electricity systems under the proposed mechanism in a realistic energy system, and the impact on the financial losses of different market participants. The simulation of the sequential operation of the heat and electricity systems over 1 year shows that the proposed electricity-aware bid selection mechanism is able to efficiently anticipate the impact of the commitment decisions of CHPs and HPs on the electricity market. As summarized in Table \ref{table:costs2}, the proposed mechanism achieves lower heat, electricity, and overall system costs compared to the decoupled one. This is achieved by switching off certain CHPs during extended periods of low day-ahead electricity prices and when their bids are invalid. For instance, we compare the heat dispatch of CHP5 and CHP6, which are located in the same electricity market zone but differ in their techno-economic characteristics. As illustrated in Fig. \ref{fig:loss}(a) the bids of CHP6 are rejected by the electricity-aware bid selection mechanism because they are invalid, and CHP5 is used to partially cover this heat production, despite offering more expensive bids. While, for a given hour, this rejected bid may be replaced with more expensive bids, over multiple days, this electricity-aware approach yields a more cost-effective commitment and dispatch in the heat system.

Additionally, as summarized in Table \ref{table:losses2}, the decoupled mechanism is myopic to the impact of day-ahead electricity prices on the heat production costs of CHPs and HPs, leading to large financial losses. As seen in Fig. \ref{fig:loss}(b) for CHP6, the electricity-aware mechanism avoids these losses by rejecting invalid bids on high-loss hours. We observe that these losses are more pronounced in winter due to the CHP's higher heat output.

\begin{table}[ht]
\centering
	\caption{Case study 2: Heat, electricity and overall energy systems commitment and dispatch costs, in $10^6$\euro, and renewable energy utilization in the electricity system, in \% of production, as a percentage of the available production, for decoupled and electricity-aware market frameworks.}
	\label{table:costs2}
		\begin{tabular}{lcc}
			\hline
			& \textbf{Decoupled} & \textbf{Electricity-Aware} \\
			\textbf{Total cost$^*$} & $1,967$ & $1,881$ ($-4.4\%$) \\
			\textbf{Heat cost}  &  $1,360$ & $1,287$ ($-5.4\%$) \\	
			\textbf{Electricity cost} &  $608$ & $594$ ($-2.2\%$)  \\
			\textbf{Renewable utilization} & $95.9$ & $96.3$ ($+0.4\%$) \\
			\hline
	\end{tabular}
\end{table}

\begin{table}[ht]
\centering
%	\resizebox{.5\textwidth}{!}{
	\caption{Case study 2: Number of hours for which invalid bids are selected by the decoupled mechanism, and the resulting financial losses for each unit. }
	\label{table:losses2}
	\centering
		\begin{tabular}{lccccccc}
			\hline
			  & \textbf{Hours} & \textbf{Losses} & & \textbf{Hours} & \textbf{Losses} \\
     		&  & \textbf{($\bm{10^3}$ \euro)} & &  & \textbf{($\bm{10^3}$ \euro)} \\
        \textbf{IS1} & $7472$ & $-6,620$ & \textbf{CHP1} & $3989$ & $-20,364$ \\
        \textbf{IS2} & $7220$ & $-6,717$ & \textbf{CHP2} & $2638$ & $-3,081$ \\
        \textbf{IS3} & $0$ & $0$ & \textbf{CHP3} & $3192$ & $-10,668$ \\
        \textbf{IS4} & $3195$ & $-8,092$ & \textbf{CHP4} & $4784$ & $-6,683$ \\
        \textbf{IS5} & $6899$ & $-11,918$ & \textbf{CHP5} & $20$ & $-148$  \\
        \textbf{IS6} & $4135$ & $-6,739$ & \textbf{CHP6} & $3078$ & $-17,203$ \\
        \textbf{HP1} & $1920$ & $-787$ & \textbf{CHP7} & $2256$ & $-7,534$ \\
        \textbf{HP2} & $2669$ & $-2,086$ & \textbf{CHP8} & $1503$ & $-5,140$  \\
        \textbf{HP3} & $2669$ & $-348$ & \textbf{CHP9} & $3212$ & $-7,531$  \\
        \textbf{HP4} & $594$ & $0$ & \textbf{CHP10} & $3161$ & $-7,395$ \\
        \textbf{HP5} & $2146$ & $-379$ & \textbf{CHP11} & $1457$ & $-5,839$ \\
        \textbf{HP6} & $1959$ & $-83$ \\
			\hline
	\end{tabular}
\end{table}

\section{Conclusion} \label{section:6}

This paper proposes a novel electricity-aware bid format and bid selection mechanism. This mechanism improves the coordination between heat and and electricity markets by allowing CHPs and HPs to offer electricity-aware heat bids which are conditioned on day-ahead electricity prices while respecting the current sequential heat and electricity market clearing procedure. These electricity-aware bids are modeled using linear bid-validity conditions.
As a result, a tractable MILP reformulation for the proposed optimization problem is developed. Finally, the value of improving the coordination between heat and electricity systems is illustrated in two case studies. The first illustrative case study shows that the proposed electricity-aware mechanism can achieve $77.6\%$ of the \textit{value of coordination} achieved by the fully integrated mechanism while maintaining a sequential heat and electricity market framework. The second case study, based on the realistic electricity and heat systems in Denmark, shows that the proposed mechanism is able to ensure cost recovery for CHPs and HPs in the heat market while reducing the operating cost of the overall energy system by $4.5\%$ and wind curtailment by $0.4\%$ compared to a decoupled mechanism. These benefits are achieved by anticipating the impact of CHPs and HPs on electricity markets. This work allows us to harness and remunerate the flexibility of CHPs and HPs at the interface between heat and electricity systems, and to achieve an efficient and cost-effective operation of the overall energy system.

This study opens up various opportunities for future work. Firstly, the proposed bid selection mechanism coordinates the participation of CHPs and HPs across multiple isolated heat networks and market zones in a centralized way. This requires a central heat system operator to collect information from independent heat market operators, which may be challenging in practice. As a potential alternative, decomposition techniques relying on consensus-based distributed algorithms could be investigated to facilitate the application of the proposed approach in the current energy system. Recent advances in the literature have introduced performance guarantees on the application of such algorithms to large-scale MILPs \cite{vujanic2016decomposition,dvorkin2018consensus,falsone2019decentralized}. Furthermore, this work does not take into account additional energy products, such as natural gas, and how their day-ahead prices may impact the validity of heat bids. As the day-ahead gas market is cleared after the heat and electricity markets, this limitation may be accounted for by developing an \textit{electricity- and gas-aware} heat bid selection mechanism \cite{byeon2019unit}.
%This research direction would also tackle the computational burden of the proposed model.
%
%the computational burden of the proposed model can be addressed using a Benders decomposition approach, as introduced in \cite{byeon2019benders}. 
%Furthermore, the proposed model provides an \textit{optimistic} hierarchical optimization problem, i.e.,  if multiple solutions of the lower-level problems exist, the one that minimizes the upper level objective ius selected. In order to relax this restrictive assumption, a \textit{pessimistic} trilevel optimization problem could be investigated \cite{dempe2014necessary}. However, such problems are challenging to solve and advanced optimization tools must be developed. 

Secondly, as previously discussed, the proposed bid selection mechanism does not guarantee efficiency and incentive compatibility, due to the unchanged design of the sequential heat and electricity markets. In order to quantify the loss of efficiency in heat and electricity markets resulting from the exercise of market power, further agent-based analysis should be conducted. Providing a rigorous analysis of potential strategic and opportunity-cost bidding behaviors across both heat and electricity markets would provide useful insights for further market designs. Furthermore, market power may be reduced by designing new market-clearing mechanisms for the day-ahead heat and electricity markets \cite{koccyiugit2020distributionally,karaca2019core}, which would require major regulatory and organizational changes.

Thirdly, the proposed model assumes perfect information on the bids of the market participants and wind power availability in the electricity market clearing. However, such information may not be communicated by the independent market operator, due to privacy concerns. Therefore, imperfect information on the parameters of the lower-level problem may be assumed using a scenario-based stochastic programming framework, or a robust counterpart of the middle- and lower-level problems. In addition, in order to mitigate the inefficiencies related to the lack of information exchange, a privacy-preserving extension of the proposed model can be developed, based on the Privacy-Preserving Stackelberg Mechanism (PPSM) introduced in \cite{fioretto2018constrained}. The PPSM allows the follower in a Stackelberg game, e.g., the electricity market operator, to share differentially-private information, e.g. bids, with the leader, e.g. the heat system operator while ensuring near-optimal coordination of the sectors.

Finally, while this work focuses on the coordination of different energy markets in the day-ahead stage, accounting for uncertainty of energy delivery in real time is essential to ensure reliability of the overall energy system. Following the work on policy-based reserves proposed by \cite{warrington2013policy} and \cite{ratha2020affine}, the proposed bid selection mechanism could be extended to co-optimize energy and operating reserves.

\bibliographystyle{ACM-Reference-Format}
\bibliography{0-sn-references.bib}


\begin{thebibliography}{51}

%%% the Web address.

\ifx \showCODEN    \undefined \def \showCODEN     #1{\unskip}     \fi
\ifx \showISBNx    \undefined \def \showISBNx     #1{\unskip}     \fi
\ifx \showISBNxiii \undefined \def \showISBNxiii  #1{\unskip}     \fi
\ifx \showISSN     \undefined \def \showISSN      #1{\unskip}     \fi
\ifx \showLCCN     \undefined \def \showLCCN      #1{\unskip}     \fi
\ifx \shownote     \undefined \def \shownote      #1{#1}          \fi
\ifx \showarticletitle \undefined \def \showarticletitle #1{#1}   \fi
\ifx \showURL      \undefined \def \showURL       {\relax}        \fi
% invisible to TeX
\providecommand\bibfield[2]{#2}
\providecommand\bibinfo[2]{#2}
\providecommand\natexlab[1]{#1}
\providecommand\showeprint[2][]{arXiv:#2}

\bibitem[Babagheibi et~al\mbox{.}(2023)]%
        {babagheibi2023integrated}
\bibfield{author}{\bibinfo{person}{Mahsa Babagheibi}, \bibinfo{person}{Ali Sahebi}, \bibinfo{person}{Shahram Jadid}, {and} \bibinfo{person}{Ahad Kazemi}.} \bibinfo{year}{2023}\natexlab{}.
\newblock \showarticletitle{An integrated design of heat and power market for energy hubs considering the security constraints of the system}.
\newblock \bibinfo{journal}{\emph{Sustainable Cities and Society}}  \bibinfo{volume}{96} (\bibinfo{year}{2023}), \bibinfo{pages}{104616}.
\newblock


\bibitem[Bobo et~al\mbox{.}(2018)]%
        {bobo2018offering}
\bibfield{author}{\bibinfo{person}{Lucien~Ali Bobo}, \bibinfo{person}{Stefanos Delikaraoglou}, \bibinfo{person}{Niklas Vespermann}, \bibinfo{person}{Jalal Kazempour}, {and} \bibinfo{person}{Pierre Pinson}.} \bibinfo{year}{2018}\natexlab{}.
\newblock \showarticletitle{Offering Strategy of a Flexibility Aggregator in a Balancing Market Using Asymmetric Block Offers}. In \bibinfo{booktitle}{\emph{Power Systems Computation Conference (PSCC)}}. IEEE.
\newblock


\bibitem[Byeon and Van~Hentenryck(2020)]%
        {byeon2019unit}
\bibfield{author}{\bibinfo{person}{Geunyeong Byeon} {and} \bibinfo{person}{Pascal Van~Hentenryck}.} \bibinfo{year}{2020}\natexlab{}.
\newblock \showarticletitle{Unit Commitment With Gas Network Awareness}.
\newblock \bibinfo{journal}{\emph{{IEEE} Transactions on Power Systems}} \bibinfo{volume}{35}, \bibinfo{number}{2} (\bibinfo{year}{2020}), \bibinfo{pages}{1327--1339}.
\newblock


\bibitem[Cococcioni et~al\mbox{.}(2018)]%
        {cococcioni2018lexicographic}
\bibfield{author}{\bibinfo{person}{Marco Cococcioni}, \bibinfo{person}{Massimo Pappalardo}, {and} \bibinfo{person}{Yaroslav~D Sergeyev}.} \bibinfo{year}{2018}\natexlab{}.
\newblock \showarticletitle{Lexicographic multi-objective linear programming using grossone methodology: Theory and algorithm}.
\newblock \bibinfo{journal}{\emph{Appl. Math. Comput.}}  \bibinfo{volume}{318} (\bibinfo{year}{2018}), \bibinfo{pages}{298--311}.
\newblock


\bibitem[Dai et~al\mbox{.}(2018)]%
        {dai2018general}
\bibfield{author}{\bibinfo{person}{Yuanhang Dai}, \bibinfo{person}{Lei Chen}, \bibinfo{person}{Yong Min}, \bibinfo{person}{Pierluigi Mancarella}, \bibinfo{person}{Qun Chen}, \bibinfo{person}{Junhong Hao}, \bibinfo{person}{Kang Hu}, {and} \bibinfo{person}{Fei Xu}.} \bibinfo{year}{2018}\natexlab{}.
\newblock \showarticletitle{A general model for thermal energy storage in combined heat and power dispatch considering heat transfer constraints}.
\newblock \bibinfo{journal}{\emph{IEEE Transactions on Sustainable Energy}} \bibinfo{volume}{9}, \bibinfo{number}{4} (\bibinfo{year}{2018}), \bibinfo{pages}{1518--1528}.
\newblock


\bibitem[Desguers et~al\mbox{.}(2024)]%
        {desguers2024integration}
\bibfield{author}{\bibinfo{person}{Thibaut Desguers}, \bibinfo{person}{Andrew Lyden}, {and} \bibinfo{person}{Daniel Friedrich}.} \bibinfo{year}{2024}\natexlab{}.
\newblock \showarticletitle{Integration of curtailed wind into flexible electrified heating networks with demand-side response and thermal storage: Practicalities and need for market mechanisms}.
\newblock \bibinfo{journal}{\emph{Energy Conversion and Management}}  \bibinfo{volume}{304} (\bibinfo{year}{2024}), \bibinfo{pages}{118203}.
\newblock


\bibitem[Dvorkin et~al\mbox{.}(2018)]%
        {dvorkin2018consensus}
\bibfield{author}{\bibinfo{person}{Vladimir Dvorkin}, \bibinfo{person}{Jalal Kazempour}, \bibinfo{person}{Luis Baringo}, {and} \bibinfo{person}{Pierre Pinson}.} \bibinfo{year}{2018}\natexlab{}.
\newblock \showarticletitle{A consensus-{ADMM} approach for strategic generation investment in electricity markets}. In \bibinfo{booktitle}{\emph{2018 IEEE Conference on Decision and Control (CDC)}}. IEEE, \bibinfo{pages}{780--785}.
\newblock


\bibitem[Energinet(2020)]%
        {energinet}
\bibfield{author}{\bibinfo{person}{Energinet}.} \bibinfo{year}{2020}\natexlab{}.
\newblock \bibinfo{booktitle}{\emph{Danish system operator, online data}}.
\newblock
\newblock
\shownote{\url{https://en.energinet.dk/}}.


\bibitem[{ETSO}(2000)]%
        {ETSO2000}
\bibfield{author}{\bibinfo{person}{{ETSO}}.} \bibinfo{year}{2000}\natexlab{}.
\newblock \bibinfo{title}{Net Transfer Capacities and Available Transfer Capacities in the Internal Market of Electricity in {E}urope}.
\newblock
\newblock
\shownote{\url{https://www.entsoe.eu/fileadmin/user_upload/_library/ntc/entsoe_NTCusersInformation.pdf}}.


\bibitem[Falsone et~al\mbox{.}(2019)]%
        {falsone2019decentralized}
\bibfield{author}{\bibinfo{person}{Alessandro Falsone}, \bibinfo{person}{Kostas Margellos}, {and} \bibinfo{person}{Maria Prandini}.} \bibinfo{year}{2019}\natexlab{}.
\newblock \showarticletitle{A decentralized approach to multi-agent {MILPs}: {F}inite-time feasibility and performance guarantees}.
\newblock \bibinfo{journal}{\emph{Automatica}}  \bibinfo{volume}{103} (\bibinfo{year}{2019}), \bibinfo{pages}{141--150}.
\newblock


\bibitem[{{FERC} and {NERC}}(2011b)]%
        {FERCreport}
\bibfield{author}{\bibinfo{person}{{{FERC} and {NERC}}}.} \bibinfo{year}{Aug. 2011}\natexlab{b}.
\newblock \bibinfo{title}{Staff Report on Outages and Curtailments During the Southwest Cold Weather Event of February 1-5, 2011: Causes and Recommendations}.
\newblock
\newblock
\shownote{\url{https://www.nerc.com/pa/rrm/ea/Pages/September-2011-Southwest-Blackout-Event.aspx}}.


\bibitem[{{FERC} and {NERC}}(2011a)]%
        {FERCactions}
\bibfield{author}{\bibinfo{person}{{{FERC} and {NERC}}}.} \bibinfo{year}{Nov. 9, 2011}\natexlab{a}.
\newblock \bibinfo{title}{Follow-up Actions in Response to August 16, 2011 Joint Report, Docket No. AD11-9-000 (delegated letter order)}.
\newblock


\bibitem[Fernqvist et~al\mbox{.}(2023)]%
        {fernqvist2023district}
\bibfield{author}{\bibinfo{person}{Niklas Fernqvist}, \bibinfo{person}{Sarah Broberg}, \bibinfo{person}{Johan Tor{\'e}n}, {and} \bibinfo{person}{Inger-Lise Svensson}.} \bibinfo{year}{2023}\natexlab{}.
\newblock \showarticletitle{District heating as a flexibility service: Challenges in sector coupling for increased solar and wind power production in Sweden}.
\newblock \bibinfo{journal}{\emph{Energy Policy}}  \bibinfo{volume}{172} (\bibinfo{year}{2023}), \bibinfo{pages}{113332}.
\newblock


\bibitem[Fioretto et~al\mbox{.}(2019)]%
        {fioretto2019ppsm}
\bibfield{author}{\bibinfo{person}{Ferdinando Fioretto}, \bibinfo{person}{Lesia Mitridati}, {and} \bibinfo{person}{Pascal Van~Hentenryck}.} \bibinfo{year}{2019}\natexlab{}.
\newblock \showarticletitle{PPSM: A Privacy-Preserving Stackelberg Mechanism}.
\newblock \bibinfo{journal}{\emph{Unknown Journal}} (\bibinfo{year}{2019}).
\newblock


\bibitem[Fioretto and Van~Hentenryck(2018)]%
        {fioretto2018constrained}
\bibfield{author}{\bibinfo{person}{Ferdinando Fioretto} {and} \bibinfo{person}{Pascal Van~Hentenryck}.} \bibinfo{year}{2018}\natexlab{}.
\newblock \showarticletitle{Constrained-based differential privacy: Releasing optimal power flow benchmarks privately}. In \bibinfo{booktitle}{\emph{International Conference on the Integration of Constraint Programming, Artificial Intelligence, and Operations Research}}. Springer, \bibinfo{pages}{215--231}.
\newblock


\bibitem[Hobbs et~al\mbox{.}(2004)]%
        {hobbs2003complementarity}
\bibfield{author}{\bibinfo{person}{Benjamin~F Hobbs}, \bibinfo{person}{Udi Helman}, {and} \bibinfo{person}{Derek~W Bunn}.} \bibinfo{year}{2004}\natexlab{}.
\newblock \showarticletitle{Complementarity-based equilibrium modeling for electric power markets}.
\newblock In \bibinfo{booktitle}{\emph{Modeling Prices in Competitive Electricity Markets}}. \bibinfo{publisher}{Wiley Series in Financial Economics}, \bibinfo{pages}{69--94}.
\newblock


\bibitem[Karaca and Kamgarpour(2019)]%
        {karaca2019core}
\bibfield{author}{\bibinfo{person}{Orcun Karaca} {and} \bibinfo{person}{Maryam Kamgarpour}.} \bibinfo{year}{2019}\natexlab{}.
\newblock \showarticletitle{Core-selecting mechanisms in electricity markets}.
\newblock \bibinfo{journal}{\emph{IEEE Transactions on Smart Grid}} \bibinfo{volume}{11}, \bibinfo{number}{3} (\bibinfo{year}{2019}), \bibinfo{pages}{2604--2614}.
\newblock


\bibitem[Kazempour et~al\mbox{.}(2018)]%
        {kazempour2018stochastic}
\bibfield{author}{\bibinfo{person}{Jalal Kazempour}, \bibinfo{person}{Pierre Pinson}, {and} \bibinfo{person}{Benjamin~F Hobbs}.} \bibinfo{year}{2018}\natexlab{}.
\newblock \showarticletitle{A Stochastic Market Design With Revenue Adequacy and Cost Recovery by Scenario: Benefits and Costs}.
\newblock \bibinfo{journal}{\emph{IEEE Transactions on Power Systems}} \bibinfo{volume}{33}, \bibinfo{number}{4} (\bibinfo{year}{2018}), \bibinfo{pages}{3531--3545}.
\newblock


\bibitem[Ko{\c{c}}yi{\u{g}}it et~al\mbox{.}(2020)]%
        {koccyiugit2020distributionally}
\bibfield{author}{\bibinfo{person}{{\c{C}}a{\u{g}}{\i}l Ko{\c{c}}yi{\u{g}}it}, \bibinfo{person}{Garud Iyengar}, \bibinfo{person}{Daniel Kuhn}, {and} \bibinfo{person}{Wolfram Wiesemann}.} \bibinfo{year}{2020}\natexlab{}.
\newblock \showarticletitle{Distributionally robust mechanism design}.
\newblock \bibinfo{journal}{\emph{Management Science}} \bibinfo{volume}{66}, \bibinfo{number}{1} (\bibinfo{year}{2020}), \bibinfo{pages}{159--189}.
\newblock


\bibitem[Lahdelma and Hakonen(2003)]%
        {lahdelma2003efficient}
\bibfield{author}{\bibinfo{person}{Risto Lahdelma} {and} \bibinfo{person}{Henri Hakonen}.} \bibinfo{year}{2003}\natexlab{}.
\newblock \showarticletitle{An efficient linear programming algorithm for combined heat and power production}.
\newblock \bibinfo{journal}{\emph{European Journal of Operational Research}} \bibinfo{volume}{148}, \bibinfo{number}{1} (\bibinfo{year}{2003}), \bibinfo{pages}{141--151}.
\newblock


\bibitem[Li et~al\mbox{.}(2016)]%
        {li2016combined}
\bibfield{author}{\bibinfo{person}{Zhigang Li}, \bibinfo{person}{Wenchuan Wu}, \bibinfo{person}{Mohammad Shahidehpour}, \bibinfo{person}{Jianhui Wang}, {and} \bibinfo{person}{Boming Zhang}.} \bibinfo{year}{2016}\natexlab{}.
\newblock \showarticletitle{Combined Heat and Power Dispatch Considering Pipeline Energy Storage of District Heating Network}.
\newblock \bibinfo{journal}{\emph{IEEE Transactions on Sustainable Energy}} \bibinfo{volume}{7}, \bibinfo{number}{1} (\bibinfo{year}{2016}), \bibinfo{pages}{12--22}.
\newblock


\bibitem[Liu et~al\mbox{.}(2015)]%
        {liu2015extending}
\bibfield{author}{\bibinfo{person}{Yanchao Liu}, \bibinfo{person}{Jesse~T Holzer}, {and} \bibinfo{person}{Michael~C Ferris}.} \bibinfo{year}{2015}\natexlab{}.
\newblock \showarticletitle{Extending the bidding format to promote demand response}.
\newblock \bibinfo{journal}{\emph{Energy Policy}}  \bibinfo{volume}{86} (\bibinfo{year}{2015}), \bibinfo{pages}{82--92}.
\newblock


\bibitem[Lund(2007)]%
        {lund2007renewable}
\bibfield{author}{\bibinfo{person}{Henrik Lund}.} \bibinfo{year}{2007}\natexlab{}.
\newblock \showarticletitle{Renewable energy strategies for sustainable development}.
\newblock \bibinfo{journal}{\emph{Energy}} \bibinfo{volume}{32}, \bibinfo{number}{6} (\bibinfo{year}{2007}), \bibinfo{pages}{912--919}.
\newblock


\bibitem[Lund et~al\mbox{.}(2010)]%
        {lund2010role}
\bibfield{author}{\bibinfo{person}{Henrik Lund}, \bibinfo{person}{Bernd M{\"o}ller}, \bibinfo{person}{Brian~Vad Mathiesen}, {and} \bibinfo{person}{A Dyrelund}.} \bibinfo{year}{2010}\natexlab{}.
\newblock \showarticletitle{The role of district heating in future renewable energy systems}.
\newblock \bibinfo{journal}{\emph{Energy}} \bibinfo{volume}{35}, \bibinfo{number}{3} (\bibinfo{year}{2010}), \bibinfo{pages}{1381--1390}.
\newblock


\bibitem[Lund et~al\mbox{.}(2014)]%
        {lund20144th}
\bibfield{author}{\bibinfo{person}{Henrik Lund}, \bibinfo{person}{Sven Werner}, \bibinfo{person}{Robin Wiltshire}, \bibinfo{person}{Svend Svendsen}, \bibinfo{person}{Jan~Eric Thorsen}, \bibinfo{person}{Frede Hvelplund}, {and} \bibinfo{person}{Brian~Vad Mathiesen}.} \bibinfo{year}{2014}\natexlab{}.
\newblock \showarticletitle{4th Generation District Heating ({4GDH}): Integrating smart thermal grids into future sustainable energy systems}.
\newblock \bibinfo{journal}{\emph{Energy}}  \bibinfo{volume}{68} (\bibinfo{year}{2014}), \bibinfo{pages}{1--11}.
\newblock


\bibitem[Madsen(2015)]%
        {Madsen2015}
\bibfield{author}{\bibinfo{person}{H. Madsen}.} \bibinfo{year}{2015}\natexlab{}.
\newblock \bibinfo{title}{Time series analysis. Course notes}.
\newblock
\newblock
\shownote{\url{http://www.imm.dtu.dk/~hmad/time.series.analysis/assignments/index.html}}.


\bibitem[Maurer et~al\mbox{.}(2023)]%
        {maurer2023toward}
\bibfield{author}{\bibinfo{person}{Jona Maurer}, \bibinfo{person}{Nicolai Tschuch}, \bibinfo{person}{Stefan Krebs}, \bibinfo{person}{Kankar Bhattacharya}, \bibinfo{person}{Claudio Ca{\~n}izares}, {and} \bibinfo{person}{S{\"o}ren Hohmann}.} \bibinfo{year}{2023}\natexlab{}.
\newblock \showarticletitle{Toward transactive control of coupled electric power and district heating networks}.
\newblock \bibinfo{journal}{\emph{Applied Energy}}  \bibinfo{volume}{332} (\bibinfo{year}{2023}), \bibinfo{pages}{120460}.
\newblock


\bibitem[McCormick(1976)]%
        {mccormick1976computability}
\bibfield{author}{\bibinfo{person}{Garth~P McCormick}.} \bibinfo{year}{1976}\natexlab{}.
\newblock \showarticletitle{Computability of global solutions to factorable nonconvex programs: {Part I—Convex} underestimating problems}.
\newblock \bibinfo{journal}{\emph{Mathematical Programming}} \bibinfo{volume}{10}, \bibinfo{number}{1} (\bibinfo{year}{1976}), \bibinfo{pages}{147--175}.
\newblock


\bibitem[Meeus et~al\mbox{.}(2005)]%
        {meeus2005development}
\bibfield{author}{\bibinfo{person}{Leonardo Meeus}, \bibinfo{person}{Konrad Purchala}, {and} \bibinfo{person}{Ronnie Belmans}.} \bibinfo{year}{2005}\natexlab{}.
\newblock \showarticletitle{Development of the internal electricity market in {E}urope}.
\newblock \bibinfo{journal}{\emph{The Electricity Journal}} \bibinfo{volume}{18}, \bibinfo{number}{6} (\bibinfo{year}{2005}), \bibinfo{pages}{25--35}.
\newblock


\bibitem[Meibom et~al\mbox{.}(2013)]%
        {meibom2013energy}
\bibfield{author}{\bibinfo{person}{Peter Meibom}, \bibinfo{person}{Klaus~Baggesen Hilger}, \bibinfo{person}{Henrik Madsen}, {and} \bibinfo{person}{Dorthe Vinther}.} \bibinfo{year}{2013}\natexlab{}.
\newblock \showarticletitle{Energy Comes Together in {D}enmark: The Key to a Future Fossil-Free {D}anish Power System}.
\newblock \bibinfo{journal}{\emph{IEEE Power \& Energy Magazine}} \bibinfo{volume}{11}, \bibinfo{number}{5} (\bibinfo{year}{2013}), \bibinfo{pages}{46--55}.
\newblock


\bibitem[Mitridati et~al\mbox{.}(2020)]%
        {mitridati2020heat}
\bibfield{author}{\bibinfo{person}{L Mitridati}, \bibinfo{person}{J. Kazempour}, {and} \bibinfo{person}{P. Pinson}.} \bibinfo{year}{2020}\natexlab{}.
\newblock \showarticletitle{Heat and electricity market coordination: A scalable complementarity approach}.
\newblock \bibinfo{journal}{\emph{European Journal of Operational Research}} \bibinfo{volume}{283}, \bibinfo{number}{3} (\bibinfo{year}{2020}), \bibinfo{pages}{1107--1123}.
\newblock


\bibitem[Mitridati and Pinson(2016)]%
        {mitridati2016optimal}
\bibfield{author}{\bibinfo{person}{Lesia Mitridati} {and} \bibinfo{person}{Pierre Pinson}.} \bibinfo{year}{2016}\natexlab{}.
\newblock \showarticletitle{Optimal coupling of heat and electricity systems: A stochastic hierarchical approach}. In \bibinfo{booktitle}{\emph{16th International Conference on Probabilistic Methods Applied to Power Systems (PMAPS)}}. \bibinfo{address}{Beijing, China}.
\newblock


\bibitem[Mitridati et~al\mbox{.}(2022)]%
        {mitridati2022differentially}
\bibfield{author}{\bibinfo{person}{Lesia Mitridati}, \bibinfo{person}{Emma Romei}, \bibinfo{person}{Gabriela Hug}, {and} \bibinfo{person}{Ferdinando Fioretto}.} \bibinfo{year}{2022}\natexlab{}.
\newblock \showarticletitle{Differentially-private heat and electricity markets coordination}. In \bibinfo{booktitle}{\emph{2022 17th International Conference on Probabilistic Methods Applied to Power Systems (PMAPS)}}. IEEE, \bibinfo{pages}{1--6}.
\newblock


\bibitem[Mitridati and Taylor(2018)]%
        {mitridati2018power}
\bibfield{author}{\bibinfo{person}{Lesia Mitridati} {and} \bibinfo{person}{Josh~A Taylor}.} \bibinfo{year}{2018}\natexlab{}.
\newblock \showarticletitle{Power Systems Flexibility from District Heating Networks}. In \bibinfo{booktitle}{\emph{Power Systems Computation Conference (PSCC)}}. \bibinfo{address}{Dublin, Ireland}.
\newblock


\bibitem[Myerson(1981)]%
        {myerson1981optimal}
\bibfield{author}{\bibinfo{person}{Roger~B Myerson}.} \bibinfo{year}{1981}\natexlab{}.
\newblock \showarticletitle{Optimal auction design}.
\newblock \bibinfo{journal}{\emph{Mathematics of operations research}} \bibinfo{volume}{6}, \bibinfo{number}{1} (\bibinfo{year}{1981}), \bibinfo{pages}{58--73}.
\newblock


\bibitem[Ordoudis et~al\mbox{.}(2016)]%
        {ordoudis2016updated}
\bibfield{author}{\bibinfo{person}{Christos Ordoudis}, \bibinfo{person}{Pierre Pinson}, \bibinfo{person}{Juan~M Morales}, {and} \bibinfo{person}{Marco Zugno}.} \bibinfo{year}{2016}\natexlab{}.
\newblock \bibinfo{title}{An updated version of the {IEEE} {RTS} 24-bus system for electricity market and power system operation studies - {DTU} Working Paper (available online)}.
\newblock
\newblock
\shownote{\url{http://orbit.dtu.dk/files/120568114/An}}.


\bibitem[Pineda and Morales(2016)]%
        {pineda2016capacity}
\bibfield{author}{\bibinfo{person}{Salvador Pineda} {and} \bibinfo{person}{Juan~M Morales}.} \bibinfo{year}{2016}\natexlab{}.
\newblock \showarticletitle{Capacity expansion of stochastic power generation under two-stage electricity markets}.
\newblock \bibinfo{journal}{\emph{Computers \& Operations Research}}  \bibinfo{volume}{70} (\bibinfo{year}{2016}), \bibinfo{pages}{101--114}.
\newblock


\bibitem[Pinson et~al\mbox{.}(2017)]%
        {pinson2017towards}
\bibfield{author}{\bibinfo{person}{Pierre Pinson}, \bibinfo{person}{Lesia Mitridati}, \bibinfo{person}{Christos Ordoudis}, {and} \bibinfo{person}{Jacob Ostergaard}.} \bibinfo{year}{2017}\natexlab{}.
\newblock \showarticletitle{Towards fully renewable energy systems: Experience and trends in Denmark}.
\newblock \bibinfo{journal}{\emph{CSEE journal of power and energy systems}} \bibinfo{volume}{3}, \bibinfo{number}{1} (\bibinfo{year}{2017}), \bibinfo{pages}{26--35}.
\newblock


\bibitem[Ratha et~al\mbox{.}(2020)]%
        {ratha2020affine}
\bibfield{author}{\bibinfo{person}{Anubhav Ratha}, \bibinfo{person}{Anna Schwele}, \bibinfo{person}{Jalal Kazempour}, \bibinfo{person}{Pierre Pinson}, \bibinfo{person}{Shahab~Shariat Torbaghan}, {and} \bibinfo{person}{Ana Virag}.} \bibinfo{year}{2020}\natexlab{}.
\newblock \showarticletitle{Affine Policies for Flexibility Provision by Natural Gas Networks to Power Systems}. In \bibinfo{booktitle}{\emph{XXI Power Systems Computation Conference}}.
\newblock


\bibitem[Sherali and Soyster(1983)]%
        {sherali1983preemptive}
\bibfield{author}{\bibinfo{person}{Hanif~D Sherali} {and} \bibinfo{person}{Allen~L Soyster}.} \bibinfo{year}{1983}\natexlab{}.
\newblock \showarticletitle{Preemptive and nonpreemptive multi-objective programming: Relationship and counterexamples}.
\newblock \bibinfo{journal}{\emph{Journal of Optimization Theory and Applications}}  \bibinfo{volume}{39} (\bibinfo{year}{1983}), \bibinfo{pages}{173--186}.
\newblock


\bibitem[Song et~al\mbox{.}(2023)]%
        {song2023potentials}
\bibfield{author}{\bibinfo{person}{Ruihao Song}, \bibinfo{person}{Thomas Hamacher}, \bibinfo{person}{Vladimir Terzija}, {and} \bibinfo{person}{Vedran~S Peri{\'c}}.} \bibinfo{year}{2023}\natexlab{}.
\newblock \showarticletitle{Potentials of using electric-thermal sector coupling for frequency control: A review}.
\newblock \bibinfo{journal}{\emph{International Journal of Electrical Power \& Energy Systems}}  \bibinfo{volume}{151} (\bibinfo{year}{2023}), \bibinfo{pages}{109194}.
\newblock


\bibitem[Sorkn{\ae}s et~al\mbox{.}(2020)]%
        {sorknaes2020smart}
\bibfield{author}{\bibinfo{person}{Peter Sorkn{\ae}s}, \bibinfo{person}{Henrik Lund}, \bibinfo{person}{IR Skov}, \bibinfo{person}{S{\o}ren Dj{\o}rup}, \bibinfo{person}{Klaus Skytte}, \bibinfo{person}{Poul~Erik Morthorst}, {and} \bibinfo{person}{Felipe Fausto}.} \bibinfo{year}{2020}\natexlab{}.
\newblock \showarticletitle{Smart Energy Markets-Future electricity, gas and heating markets}.
\newblock \bibinfo{journal}{\emph{Renewable and Sustainable Energy Reviews}}  \bibinfo{volume}{119} (\bibinfo{year}{2020}), \bibinfo{pages}{109655}.
\newblock


\bibitem[Tosatto and Chatzivasileiadis(2020)]%
        {tosatto2019hvdc}
\bibfield{author}{\bibinfo{person}{Andrea Tosatto} {and} \bibinfo{person}{Spyros Chatzivasileiadis}.} \bibinfo{year}{2020}\natexlab{}.
\newblock \showarticletitle{{HVDC} loss factors in the {N}ordic Market}. In \bibinfo{booktitle}{\emph{Power Systems Computation Conference (PSCC)}}. \bibinfo{address}{Porto, Portugal}.
\newblock


\bibitem[Varmelast(2018)]%
        {Varmelast}
\bibfield{author}{\bibinfo{person}{Varmelast}.} \bibinfo{year}{2018}\natexlab{}.
\newblock \bibinfo{title}{Heating Plans and Day-Ahead Heat Market Clearing}.
\newblock
\newblock
\shownote{\url{http://varmelast.dk/en/heat-plans/heating-plans}}.


\bibitem[Virasjoki et~al\mbox{.}(2018)]%
        {virasjoki2018market}
\bibfield{author}{\bibinfo{person}{Vilma Virasjoki}, \bibinfo{person}{Afzal~Saeed Siddiqui}, \bibinfo{person}{Behnam Zakeri}, {and} \bibinfo{person}{Ahti Salo}.} \bibinfo{year}{2018}\natexlab{}.
\newblock \showarticletitle{Market Power with Combined Heat and Power Production in the {N}ordic Energy System}.
\newblock \bibinfo{journal}{\emph{IEEE Transactions on Power Systems}} \bibinfo{volume}{33}, \bibinfo{number}{5} (\bibinfo{year}{2018}), \bibinfo{pages}{5263--5275}.
\newblock


\bibitem[Vujanic et~al\mbox{.}(2016)]%
        {vujanic2016decomposition}
\bibfield{author}{\bibinfo{person}{Robin Vujanic}, \bibinfo{person}{Peyman~Mohajerin Esfahani}, \bibinfo{person}{Paul~J Goulart}, \bibinfo{person}{S{\'e}bastien Mari{\'e}thoz}, {and} \bibinfo{person}{Manfred Morari}.} \bibinfo{year}{2016}\natexlab{}.
\newblock \showarticletitle{A decomposition method for large scale {MILPs}, with performance guarantees and a power system application}.
\newblock \bibinfo{journal}{\emph{Automatica}}  \bibinfo{volume}{67} (\bibinfo{year}{2016}), \bibinfo{pages}{144--156}.
\newblock


\bibitem[Wang et~al\mbox{.}(2010)]%
        {wang2010new}
\bibfield{author}{\bibinfo{person}{J Wang}, \bibinfo{person}{S Kennedy}, {and} \bibinfo{person}{J Kirtley}.} \bibinfo{year}{2010}\natexlab{}.
\newblock \showarticletitle{A new wholesale bidding mechanism for enhanced demand response in smart grids}. In \bibinfo{booktitle}{\emph{2010 Innovative Smart Grid Technologies (ISGT)}}. IEEE, \bibinfo{pages}{1--8}.
\newblock


\bibitem[Warrington et~al\mbox{.}(2013)]%
        {warrington2013policy}
\bibfield{author}{\bibinfo{person}{Joseph Warrington}, \bibinfo{person}{Paul Goulart}, \bibinfo{person}{S{\'e}bastien Mari{\'e}thoz}, {and} \bibinfo{person}{Manfred Morari}.} \bibinfo{year}{2013}\natexlab{}.
\newblock \showarticletitle{Policy-based reserves for power systems}.
\newblock \bibinfo{journal}{\emph{IEEE Transactions on Power Systems}} \bibinfo{volume}{28}, \bibinfo{number}{4} (\bibinfo{year}{2013}), \bibinfo{pages}{4427--4437}.
\newblock


\bibitem[Wilson(1977)]%
        {wilson1977bidding}
\bibfield{author}{\bibinfo{person}{Robert Wilson}.} \bibinfo{year}{1977}\natexlab{}.
\newblock \showarticletitle{A bidding model of perfect competition}.
\newblock \bibinfo{journal}{\emph{The Review of Economic Studies}} \bibinfo{volume}{44}, \bibinfo{number}{3} (\bibinfo{year}{1977}), \bibinfo{pages}{511--518}.
\newblock


\bibitem[Zheng et~al\mbox{.}(2018)]%
        {zheng2018integrated}
\bibfield{author}{\bibinfo{person}{Jinfu Zheng}, \bibinfo{person}{Zhigang Zhou}, \bibinfo{person}{Jianing Zhao}, {and} \bibinfo{person}{Jinda Wang}.} \bibinfo{year}{2018}\natexlab{}.
\newblock \showarticletitle{Integrated heat and power dispatch truly utilizing thermal inertia of district heating network for wind power integration}.
\newblock \bibinfo{journal}{\emph{Applied Energy}}  \bibinfo{volume}{211} (\bibinfo{year}{2018}), \bibinfo{pages}{865--874}.
\newblock


\bibitem[Zugno et~al\mbox{.}(2016)]%
        {zugno2016commitment}
\bibfield{author}{\bibinfo{person}{Marco Zugno}, \bibinfo{person}{Juan~Miguel Morales}, {and} \bibinfo{person}{Henrik Madsen}.} \bibinfo{year}{2016}\natexlab{}.
\newblock \showarticletitle{Commitment and dispatch of heat and power units via affinely adjustable robust optimization}.
\newblock \bibinfo{journal}{\emph{Computers \& Operations Research}}  \bibinfo{volume}{75} (\bibinfo{year}{2016}), \bibinfo{pages}{191--201}.
\newblock


\end{thebibliography}

\appendix 

\section{Proof of Proposition \ref{prop1}} \label{AppendixA}

%%%%%%
The equilibrium problem representing the sequential heat and electricity market clearing in \eqref{heat}-\eqref{elec} can be expressed in a compact manner as

\begin{small}
\begin{subequations} \label{eq}
	\begin{align}
&   \bm{x^\textbf{H}} \in \text{ primal solution of }
    \begin{Bmatrix}
        \underset{{\bm{x^\textbf{H}}  \geq \bm{0} }}{\min} \ & c^{\text{H}^\top}   \bm{x^\textbf{H}} \\
        \ \text{ s.t. } & A^{\text{H}}  	 \bm{x^\textbf{H}} + B^{\text{H}}   \bm{z} \geq b^{\text{H}} 
    \end{Bmatrix} \\
    & \bm{x^\textbf{E}} \in \text{primal solution of } \begin{Bmatrix}
        \underset{{\bm{x^\textbf{E}} \geq \bm{0}}}{\min} \ & c^{\text{E}^\top} \bm{x^\textbf{E}} \\
        \ \text{ s.t. } & A^{\text{E}}  \bm{x^\textbf{E}}   +  B^{\text{E}}  \bm{x^\textbf{H}} \geq b^{\text{E}}
    \end{Bmatrix} \label{eq1}
	\end{align}
\end{subequations}   
\end{small}
\noindent Similarly, the proposed lexicographic optimization problem \eqref{lex} is formulated in a compact manner as \eqref{ul_compact4}. This lexicographic optimization problem can be solved in two steps. Firstly, find the optimal heat dispatch cost such that

\begin{small}
    \begin{subequations} \label{lex_step1}
    \begin{alignat}{3}
        & \Theta^{\text{H}^*} = && \underset{{\bm{x^\textbf{H}} , \bm{x^\textbf{E}}  \geq \bm{0} }}{\min} \ && c^{\text{H}^\top}   \bm{x^\textbf{H}} \label{lex_step1_1} \\
        & \quad && \quad \text{s.t. } \ && \eqref{heat2} - \eqref{heat8} , \quad \eqref{elec2} - \eqref{elec8} \label{lex_step1_2}
    \end{alignat}
    \end{subequations}
\end{small}
\noindent Secondly, find an optimal heat and electricity dispatch such that:

\begin{small}
    \begin{subequations} \label{lex_step2}
    \begin{alignat}{3}
        & \{x^{\text{H}^*},x^{\text{E}^*}\} \in && \ \underset{{\bm{x^\textbf{H}} , \bm{x^\textbf{E}}  \geq \bm{0} }}{\text{argmin}} \ && c^{\text{E}^\top}   \bm{x^\textbf{E}} \label{lex_step2_1} \\
        & \quad && \ \text{ s.t. } \ && \eqref{heat2} - \eqref{heat8} , \quad \eqref{elec2} - \eqref{elec8} \label{lex_step2_2} \\
        & \quad && \quad && c^{\text{H}^\top}   \bm{x^\textbf{H}}  \leq \Theta^{\text{H}^*} \label{lex_step2_3}
    \end{alignat}
    \end{subequations}
\end{small}
\noindent Due to Constraints \eqref{lex_step2_2} and \eqref{lex_step2_3}, any optimal solution $\{x^{\text{H}^*},x^{\text{E}^*}\}$ to \eqref{lex_step1}-\eqref{lex_step2}, is also optimal to \eqref{eq1}.

\section{Proof of Proposition \ref{prop2}} \label{AppendixB}

%%%%%%%%%%%%
We consider the following approximation of the lexicographic optimization problem \eqref{ul_compact4}, with $\gamma \in ]0,1[$:

\begin{small}
\begin{subequations} \label{lp_gamma}
	\begin{alignat}{7}
	&   \underset{\bm{x^\textbf{H}} , \bm{x^\textbf{E}} \geq \bm{0} }{\min} \ && \gamma c^{\text{H}^\top} \bm{x^\textbf{H}} + \left(1-\gamma\right) c^{\text{E}^\top}   \bm{x^\textbf{E}}  \label{lp_gamma1} \\
	&   \quad   \text{s.t.}  &&  A^\text{H}     \bm{x^\textbf{H}} + B^\text{H}  \bm{z}  \geq b^\text{H} \label{lp_gamma2} \\
	& \quad          &&   A^\text{E}  \bm{x^\textbf{E}}   +  B^\text{E}  \bm{x^\textbf{H}} \geq b^\text{E}, \label{lp_gamma3}
	\end{alignat}
\end{subequations}
\end{small}
\noindent where $\bm{y^\textbf{E}} $ is obtained as the dual variable associated with constraint \eqref{lp_gamma3} \cite{byeon2019unit}. As a result, problem \eqref{ul} can be approximated by the following linear bilevel optimization problem:

\begin{small}
\begin{subequations} \label{ul_gamma}
	\begin{alignat}{7}
	& \min_{ \overset{\bm{z}\in\{0,1\}^\text{N},\bm{x^\textbf{H}} , \bm{x^\textbf{E}} \geq \bm{0} }{\underset{  \bm{y^\textbf{H}},\bm{y}^\textbf{E} \geq \bm{0}}{}} }  && \gamma c^{0^\top}  \bm{z}  + \gamma c^{\text{H}^\top}   \bm{x^\textbf{H}} + \left( 1-\gamma \right) c^{\text{E}^\top}  \bm{x^\textbf{E}} \label{ul_gamma1} \\
	& \quad \quad \text{s.t.} &&  \bm{z} \in \mathcal{Z}^{\text{bid}} \label{ul_gamma2} \\
	& \quad &&  A^{\text{bid}}  \bm{z}  +     \dfrac{1}{(1-\gamma)}  B^{\text{bid}} \bm{y^\textbf{E}}   \geq  b^{\text{bid}} \label{ul_gamma3} \\
	& \quad &&   \{ \bm{x^\textbf{H}} , \bm{y^\textbf{E}} \} \in \text{ primal and dual sol. of }  \eqref{lp_gamma}. \label{ul_gamma4}
	\end{alignat}
\end{subequations}
\end{small}
Besides, by strong duality of the lower-level problem \eqref{ul_gamma4}, problem \eqref{ul_gamma} is equivalent to \eqref{primal_dual}.

It remains to show that problem \eqref{primal_dual} is an asymptotic approximation to problem \eqref{ul_compact}, i.e.,  as $\gamma \rightarrow 1$ the solutions to problem \eqref{primal_dual} become optimal solutions to problem \eqref{ul_compact}. By introducing the auxiliary variables $\bm{\tilde{y}^\textbf{H}}=\dfrac{\bm{y^\text{H}}}{\gamma}$, and $\bm{\tilde{y}^\textbf{E}}=\dfrac{\bm{y^\text{E}}}{1 - \gamma}$, problem \eqref{primal_dual} is equivalent to:

\begin{small}
\begin{subequations} \label{primal_dual_switch}
\begin{alignat}{2}
& \min_{\overset{\bm{z}\in\{0,1\}^N,\bm{x^\textbf{H}} \geq \bm{0} }{\underset{\bm{x^\textbf{E}} \geq \bm{0} , \bm{y^\textbf{H}},\bm{y}^\textbf{E}}{}}} &&  \gamma c^{0^\top}  \bm{z}  + \gamma c^{\text{H}^\top}  \bm{x^\textbf{H}} + \left( 1-\gamma \right) c^{\text{E}^\top}  \bm{x^\textbf{E}} \label{primal_dual_switch1} \\
& \quad \quad \text{s.t.} &&  \bm{z} \in \mathcal{Z}^{\text{bid}} \label{primal_dual_switch2} \\
& \quad &&  A^{\text{bid}}  \bm{z}  + B^{\text{bid}} \bm{\tilde{y}^\textbf{E}}  \geq  b^{\text{bid}} \label{primal_dual_switch3} \\
& \quad  && A^{\text{H}}     \bm{x^\textbf{H}} + B^{\text{H}}    \bm{z} \geq b^{\text{H}} \label{primal_dual_switch4} \\
 &  \quad    &&  A^{\text{E}}  \bm{x^\textbf{E}}   +  B^{\text{E}}  \bm{x^\textbf{H}} \geq b^{\text{E}} \label{primal_dual_switch5} \\
& \quad       && \bm{\tilde{y}^{\textbf{H}^\top}}  A^{\text{H}}   + \dfrac{\left( 1-\gamma \right)}{\gamma}\bm{\tilde{y}^{\textbf{E}^\top}} B^{\text{E}}   \leq c^{\text{H}^\top} \label{primal_dual_switch6} \\
 &  \quad    &&  \bm{\tilde{y}^{\textbf{E}^\top}} A^{\text{E}}  \leq c^{\text{E}^\top} \label{primal_dual_switch7} \\
 & \quad && \bm{\tilde{y}^{\textbf{H}^\top}}   \left( b^{\text{H}} - B^{\text{H}}    \bm{z}  \right) - c^{\text{H}^\top} \bm{x^\textbf{H}} \geq  \dfrac{\left( 1-\gamma \right)}{\gamma} \left( c^{\text{E}^\top} \bm{x^\textbf{E}} - \bm{\tilde{y}^{\textbf{E}^\top}} b^{\text{E}} \right).  \label{primal_dual_switch8} 
\end{alignat}
\end{subequations}
\end{small}
For the value of the unit commitment variable $\bm{z}$ fixed to $z^*$, let us denote $\Theta(z^*)$ the optimal objective value to \eqref{ul_compact}, and $\tilde{\Theta}(z^*)$ and $\{x^{\text{H}^*},x^{\text{E}^*},y^{\text{H}^*},y^{\text{E}^*}\}$ the optimal objective and solutions to \eqref{primal_dual_switch}. As $\gamma \rightarrow 1$,  \eqref{primal_dual_switch6} and \eqref{primal_dual_switch8} become 

\begin{small}
   \begin{subequations} \label{primal_dual_switch_new}
\begin{alignat}{1}
&  \bm{\tilde{y}^{\textbf{H}^\top}}  A^{\text{H}}  \leq c^{\text{H}^\top} \label{primal_dual_switch_new7} \\
 &  \bm{\tilde{y}^{\textbf{H}^\top}}   \left( b^{\text{H}} - B^{\text{H}}    \bm{z}  \right)  \geq  c^{\text{H}^\top} \bm{x^\textbf{H}}.  \label{primal_dual_switch_new8} 
\end{alignat}
\end{subequations}
\end{small}
\noindent Constraint \eqref{primal_dual_switch4} guarantees that $x^{\text{H}^*}$ is feasible to problem \eqref{lex_step1} with $\bm{z}$ fixed to $z^*$. Additionally,  \eqref{primal_dual_switch_new7} guarantees that $y^{\text{H}^*}$ becomes feasible to the dual of problem \eqref{lex_step1} with $\bm{z}$ fixed to $z^*$ when $\gamma \rightarrow 1$. Moreover,  \eqref{primal_dual_switch_new8} guarantees that $x^{\text{H}^*}$ and $y^{\text{H}^*}$, together, satisfy the strong duality equation of problem \eqref{lex_step1} with $\bm{z}$ fixed to $z^*$ when $\gamma \rightarrow 1$.
Therefore, $x^{\text{H}^*}$ and $y^{\text{H}^*}$ approximate a primal and dual optimal solution to problem \eqref{lex_step1} with $\bm{z}$ fixed to $z^*$ when $\gamma \rightarrow 1$. This implies that $x^{\text{H}^*}$ and $y^{\text{H}^*}$ become feasible solutions to \eqref{ul_compact} when $\gamma \rightarrow 1$. Moreover, the combination of  \eqref{primal_dual_switch6} $ \times x^{\text{H}^*} $ and \eqref{primal_dual_switch8} gives

\begin{small}
\begin{align} \label{reformulation1}
& \bm{\tilde{y}^{\textbf{H}^\top}}   \left( b^{\text{H}} - B^{\text{H}}    \bm{z}  - A^{\text{H}}  x^{\text{H}^*} \right) \geq  \dfrac{\left( 1-\gamma \right)}{\gamma} \left( c^{\text{E}^\top} \bm{x^\textbf{E}} - \bm{\tilde{y}^{\textbf{E}^\top}} \left(b^{\text{E}} - B^{\text{E}} x^{\text{H}^*} \right) \right).
\end{align}
\end{small}
\noindent It follows from  \eqref{reformulation1} and \eqref{primal_dual_switch4} that, for any gamma $\gamma \in ]0,1[$:

\begin{small}
\begin{equation} \label{reformulation2}
\bm{\tilde{y}^{\textbf{E}^\top}} \left(b^{\text{E}} - B^{\text{E}} x^{\text{H}^*} \right) \geq  c^{\text{E}^\top} \bm{x^\textbf{E}}.
\end{equation}
\end{small}
\noindent Constraints \eqref{primal_dual_switch5} and \eqref{primal_dual_switch7} guarantee that $x^{\text{E}^*}$ and $y^{\text{E}^*}$ are feasible primal and dual solutions to problem \eqref{lex_step2} with $\bm{x^\text{H}}$ fixed to $x^{\text{H}^*}$. Additionally,  \eqref{reformulation2} guarantees that $x^{\text{E}^*}$ and $y^{\text{E}^*}$, together, satisfy the strong duality equation of problem \eqref{lex_step2}. Therefore, $x^{\text{E}^*}$ and $y^{\text{E}^*}$ are the primal and dual optimal solutions to problem \eqref{lex_step2} with $\bm{x^\text{H}}$ fixed to $x^{\text{H}^*}$ for any $\gamma \in ]0,1[$.

In summary, $x^{\text{H}^*}$ is a feasible solution to \eqref{lex_step1}, which converges towards an optimal solution when $\gamma \rightarrow 1$, and $y^{\text{E}^*}$ is an optimal dual solution of the lower-level problem for any $\gamma \in ]0,1[$. Hence, problem \eqref{primal_dual} always provides a feasible solution to problem \eqref{ul_compact}, which converges towards the optimal solution when $\gamma \rightarrow 1$.

%%%%%%%%
Besides, note that the theoretical results in \cite{sherali1983preemptive} show that there exists a finite $0< M <1$ such that the lexicographic optimization problem \eqref{ul_compact4} is equivalent to a linear program (LP) minimizing the single objective function $c^{\text{H}^\top}   \bm{x^\textbf{H}} + M c^{\text{E}^\top}   \bm{x^\textbf{E}} $. Similarly, \cite{cococcioni2018lexicographic} shows that \eqref{ul_compact4} is equivalent to an LP minimizing the single objective function $c^{\text{H}^\top}   \bm{x^\textbf{H}} + c^{\text{E}^\top} \bm{x^\textbf{E}} \text{\textcircled{1}}^{-1} $, where \textcircled{1} represents the number of elements in the set of natural numbers $\mathbb{N}$.

\end{document}